\documentclass[11pt]{article}
\pdfoutput=1
\usepackage[utf8]{inputenc}
\usepackage[T1]{fontenc}
\usepackage{float}
\usepackage{authblk}
\usepackage[normalem]{ulem}
\usepackage{color, colortbl}
\usepackage{amsmath}
\usepackage{amsfonts}
\usepackage{amsthm}
\usepackage{adjustbox}
\usepackage{graphicx}
\usepackage{mathtools}
\usepackage{fullpage}
\usepackage{todonotes}
\usepackage{subcaption}
\usepackage[mathlines]{lineno}
\usepackage[colorlinks = true,
            linkcolor = teal,
            urlcolor  = teal,
            citecolor = teal]{hyperref}
\usepackage[style=numeric,sorting=none,maxbibnames=20]{biblatex}
\newcolumntype{P}[1]{>{\centering\arraybackslash}p{#1}}

\usepackage{tikz}
\usetikzlibrary{calc}

\tikzset{
  solid node/.style={circle,draw,inner sep=1.5,fill=black},
  hollow node/.style={circle,draw,inner sep=1.5}
}

\title{Optimal design of experiments to identify latent behavioral types}
\author[1,2]{Stefano Balietti}
\author[3]{Brennan Klein}
\author[3,4,5,6]{Christoph Riedl\thanks{c.riedl@neu.edu}}
\affil[1]{Center for European Social Science Research Universität Mannheim, MZES, 68131 Mannheim, Germany}
\affil[2]{Research Center for Environmental Economics, Heidelberg University, 69117 Heidelberg, Germany}
\affil[3]{Network Science Institute, Northeastern University, Boston, MA, USA}
\affil[4]{D'Amore-McKim School of Business, Northeastern University, Boston, MA 02115, USA}
\affil[5]{Khoury College of Computer Sciences, Northeastern University, Boston, MA 02115, USA}
\affil[6]{Institute for Quantitative Social Science, Harvard University, Cambridge, MA 02138, USA}

\begin{document}
\maketitle
\pagenumbering{arabic}

\begin{abstract}
Bayesian optimal experiments that maximize the information gained from collected data are critical to efficiently identify behavioral models. We extend a seminal method for designing Bayesian optimal experiments by introducing two computational improvements that make the procedure tractable: (1) a search algorithm from artificial intelligence that efficiently explores the space of possible design parameters, and (2) a sampling procedure which evaluates each design parameter combination more efficiently. We apply our procedure to a game of imperfect information to evaluate and quantify the computational improvements. We then collect data across five different experimental designs to compare the ability of the optimal experimental design to discriminate among competing behavioral models against the experimental designs chosen by a ``wisdom of experts'' prediction experiment. We find that data from the experiment suggested by the optimal design approach requires significantly less data to distinguish behavioral models (i.e., test hypotheses) than data from the experiment suggested by experts. Substantively, we find that reinforcement learning best explains human decision-making in the imperfect information game and that behavior is not adequately described by the Bayesian Nash equilibrium. Our procedure is general and computationally efficient and can be applied to dynamically optimize online experiments.

\end{abstract}

\section{Introduction}
\label{intro}

Experimentation in the social sciences is a fundamental tool for understanding the mechanisms and heuristics that underlie human behavior. At the same time, running experiments is a costly process and requires careful design in order to test hypotheses while maximizing statistical power. This experimental design process is often guided by the intuition of the scientists conducting the research. While there are many benefits in relying on the intuition of experienced researchers, there is often a lack of principled guides when choosing which experiment to run \cite{Fisher1936, Hill1995}. As a result, experiments may have low power to distinguish between different models of behavior \cite{salmon2001evaluation} and lead to increased costs for data collection. At worst, they lead to reduced effect sizes, and incorrect rejection or acceptance of a null hypothesis \cite{berman_phacking_ab_2016}.

A growing body of researchers including social scientists, computer scientists, and industry professionals have explored ways in which experiment selection can be optimized and how artificial intelligence (AI) can be used to select experiments \cite{Rzhetsky2015ChoosingDiscovery}. For example, researchers now optimize the experiments they run using statistical techniques (e.g., Thompson sampling) to preferentially assign participants to experimental treatments in order to optimize the \textit{treatment effect}. This allows researchers to easily decide which treatment among many possible treatment arms is most effective in maximizing a certain outcome, such as the response rate in a marketing campaign \cite{Eckles2014, Letham2017,schwartz_bandit_experiments_2017,zhou2018search}. Other techniques involve optimizing the length of the experiment, the number of participants needed, or the sequence of questions in behavioral batteries \cite{Wang2010DynamicallyParameters, Imai2016EstimatingDesign, pooseh_battery_2018, chapman_DOSE2_2018}. In each case, optimizing experimental design has proved fruitful.

However, these techniques tend to focus on the optimization of individuals' decisions (e.g., finding the best sequence of survey questions for estimating an individual's risk preferences). These approaches are ill-suited for those situations involving strategic interactions, information asymmetries, or network effects \cite{david1985clio, katz1985network, bramoulle2020peer}. In these contexts, agents with heterogeneous preferences derive different utilities for outcomes that depend on the actions of others agents \cite{camerer2011behavioral, fudenberg1991game}. For example, for goods with network effects, the utility of one consumer depends not only on that consumer's preference for the product but also on how many others adopt the good \cite{tauber_people_shop_1972, gilchrist_talk_about_2016}. Attempting to optimize experiments with such strategic complements thus requires a different approach.

We focus on a different set of optimal experimental design procedures that optimize interactive experiments mapping human decision-making to \textit{behavioral types}, i.e., mathematical models that predict human behavior under specific sociological, economic, and psychological hypotheses \cite{harsanyi1967games}. By building and testing various models about how individuals of a specific behavioral type make strategic decisions, modelers and experimenters can make fine-tuned predictions about the likely behavior of participants in experimental settings. As a result, we are better  positioned to anticipate responses to societal, market, or technological changes \cite{mcintyre_network_markets_2014}. 

In this paper, we introduce an AI-based optimal design procedure that maximizes the information gained from running a behavioral experiment. Our procedure recommends which experiments to run (i.e., which parameter values to use for data collection) in order to maximally diverge the predictions of multiple competing models of human behavior. The output of this procedure takes the form of a coordinate---a point in the space of all possible experiments---that corresponds to optimal experimental parameters. By way of example, consider an experiment that measures gambling tendency and is parameterized by a maximum payout (\textit{M}) and a payout multiplier (\textit{m}). The experimenter's question is which values to choose for $M$ and $m$ for data collection. Our protocol would output a particular experimental design in the form of a coordinate, (\textit{M}, \textit{m}), for example, \textit{M}=\$4.25 and \textit{m}=1.2 (as opposed to \textit{M}=\$4.15 and \textit{m}=1.3). While knowing the optimal experimental design is valuable for \textit{collecting} informative data, \textit{finding} the optimal design is a computationally-intensive task. The approach generally requires evaluating complex models such as whether a set of observations corresponds to a Nash equilibrium. We propose two methodological innovations that build on seminal methods in Bayesian optimal design \cite{El-Gamal1996EconomicalDesign}. The first one addresses \textit{which} parameter combinations are evaluated, the second addresses \textit{how} the information content of a given parameter combination is evaluated. Together, both improvements reduce the computational costs of designing optimal experiments by several orders of magnitude while maintaining accuracy. We give a brief intuition behind our two improvements.

First, we use an adaptive search algorithm based on a Gaussian Process to efficiently search the space of all possible experimental parameter combinations \cite{Contal2013}. Absent such an improvement, the optimal design procedure has to follow a brute-force approach that evaluates all possible experimental designs. The intuition behind this improvement is that the search algorithm adaptively explores those experimental designs that are likely very good (designs with high information content) more thoroughly, while reducing the exploration of parameter combinations that are similar to those already known to be bad (low information content).

Second, we develop a sampling technique that evaluates the informativeness of a given experimental design on the basis of simulated \textit{likely} datasets instead of \textit{all observable datasets}. This allows us to reduce the size of datasets while simultaneously increasing the relevance of data points in the dataset and thus retaining accuracy despite the smaller size.

\begin{figure}[t!]
\centering
\includegraphics[width=1.0\columnwidth]{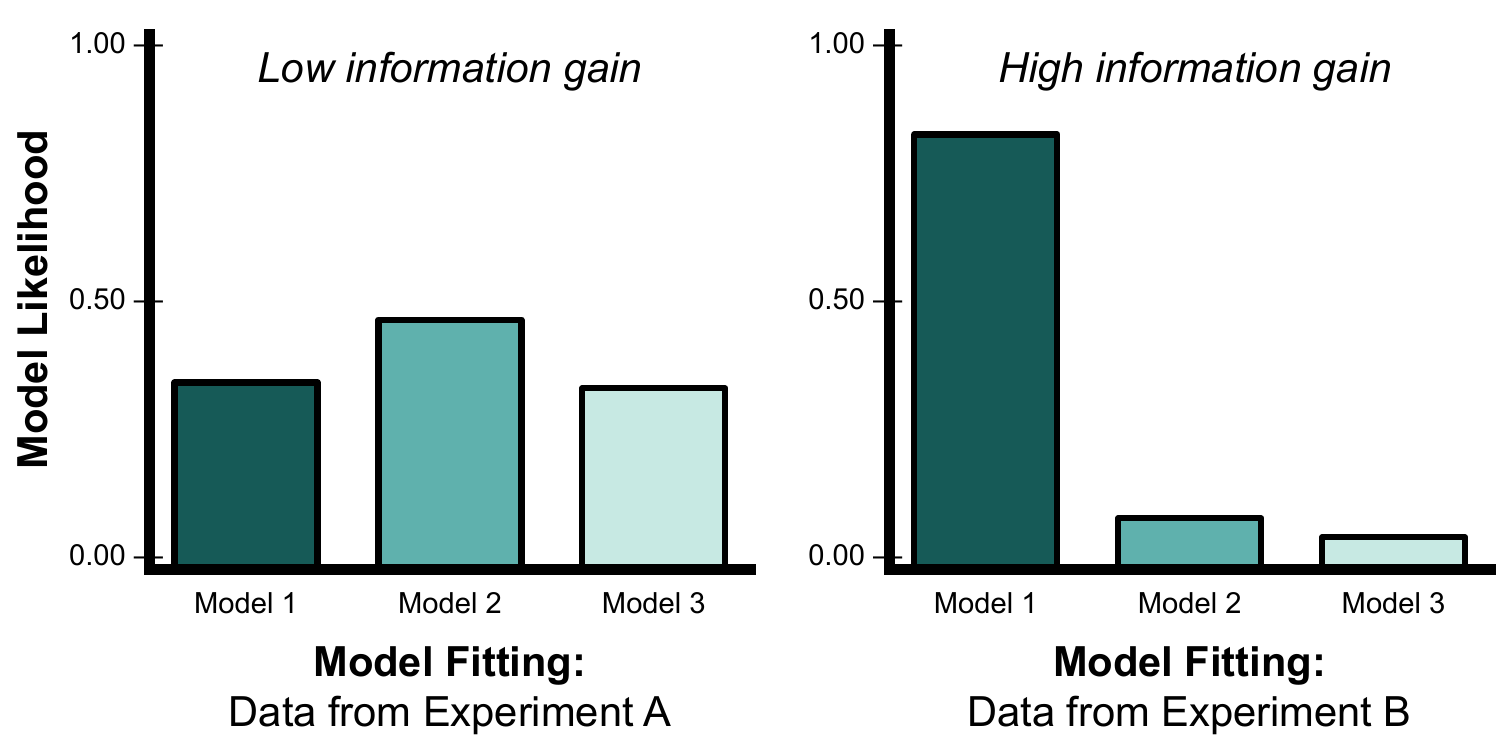}
\caption{\textbf{Illustration of lowly- vs.~highly-informative experiments}. Models 1, 2, and 3 assign likelihoods to the data generated in two hypothetical experiments: A and B. Experiment A does not allow researchers to say with certainty which model fits the data best; in other words, the design of Experiment A generates data that contains ``Low Information.'' On the other hand, Experiment B has a clear winner: Model 1 predicted the observed data as highly likely, while Models 2 and 3 did not. Therefore, the data generated by Experiment B contains ``High Information.'' The optimal experimental design procedure outlined in this paper returns values for experimental parameters that are expected to maximally distinguish the likelihoods of the competing models, as in Experiment B.}
\label{fig:sample_infogain}
\end{figure}

There are six main parts in this article. First, we describe how to optimize experimental design for information gain, and we lay out our methodological improvements to this protocol. Second, we evaluate and quantify the extent of these computational improvements. Third, we use our improved procedure to find the optimal experimental design of a classic two-person imperfect information game \cite{El-Gamal1996EconomicalDesign}. Fourth, we conduct an expert prediction experiment to see which experimental designs domain experts would recommend, and we compare these predictions to the optimal experiment suggested by our algorithmic approach. Fifth, we implement and run five experiments---each with different parameterizations---to illustrate the crucial role that information gain plays in distinguishing competing models of behavior and to showcase the adaptive nature of our approach.
Finally, we expand the set of behavioral models (with Roth-Erev reinforcement learning) to illustrate how the optimal design procedure can be used iteratively to test behavioral hypotheses as they arise.

\section{Optimal Experimental Design: Theory}

The notion of information gain is intuitive, but it is seldom defined precisely. Here, information refers to \textit{bits}---a mathematical quantity that encodes counterfactual knowledge and describes the amount of uncertainty or noise in a system. In the context of experimentation, we gain information by becoming more certain about the relationships between what we predict and what we observe in the data that we collect. This is equivalent to saying that we are becoming more certain about the models and theories we use to describe the world. Different theories generate different models that predict how humans will behave, and an informative experiment is one that is able to rule out models that cannot adequately explain the data that they are confronted with. With an experimental design that maximizes information gain, a small batch of data may be sufficient to determine which of these competing models is best suited to describe the data observed, and---importantly---this is not necessarily true for experimental designs that do not (see Fig.~\ref{fig:sample_infogain} for a visual example of experiments with low and high information gain).

Using this formalism, we can compare any different models' likelihoods of experimental outcomes based on how much relative information each distribution will provide to us. This concept has clear applications to the optimal design of experiments for testing competing hypotheses about how humans behave. All that is required is to specify competing hypotheses in the form of generative models of human behavior, assigning a certain likelihood to any given outcome of the game (the likelihoods must sum to 1). The \textit{best} experiment we can run is then the experiment that will maximally distinguish the likelihood distributions of the different competing models. This is formally achieved by maximizing the Kullback-Leibler (KL) divergence \cite{Kullback1951} between two distributions, $P$ and $Q$:
\begin{align}
D_{KL}[P||Q] &= \sum_{x \in X} p(x)\log{\frac{p(x)}{q(x)}} \hspace{0.5cm}\text{where}\hspace{0.5cm} D_{KL}[P||Q] \neq D_{KL}[Q||P]
\end{align}

Here, we refer to the \textit{optimal} experiment as the particular set of design parameters, $\theta \in \Theta$, that are predicted to maximally distinguish between \textit{n} competing generative models of behavior in a given task. As in El Gamal \& Palfrey (1996), we use a one-sided form of the KL divergence, which compares $n-1$ competing models to a single model, expressed as $I(1;\theta)$, as expressed below:
\begin{align}
I(1;\theta) = \sum_{x \in X} l_1(x;\theta) \log \Bigg(\frac{(1-p_1)l_1(x;\theta)}{
{\displaystyle \sum_{i=2}^n p_i l_i(x;\theta)}}\Bigg)
\label{eq:DKL}
\end{align}
where $l_i(x;\theta)$ is the likelihood that model $i$ assigns to observing a specific dataset $x$ in the space of all possible datasets $X$ under the experimental design $\theta$, and $p_i$ is the experimenter prior over model $i$ (both the sum of priors over models and the sum of model likelihoods over datasets are equal to 1). In other words, Equation \ref{eq:DKL} assigns a value for each combination of experimental parameters in $\theta$. This value corresponds to the amount of information we would expect to gain as a result of running an experiment with that particular combination of parameters\footnote{ It is important to note that this difference metric is a directed, asymmetric measure. Wang et al.~use a similar metric but propose using the average KL divergence \cite{Wang2010DynamicallyParameters}, which can be expressed as $I(\theta) = \displaystyle \sum_{i}^n p_i I(i;\theta)$.}. By computing this value for every coordinate in a grid comprised of each parameter combination, we are able to create an ``information surface'' where the value of each point represents the information theoretic difference between one model and $n-1$ competing models (for a visual example of this, see Fig.~\ref{fig:surface_compare}).

As an illustration, consider the following example. The three authors of this paper each bet on a different model that they believe \textit{best} describes human behavior in a given game. In order to determine who bet on the best model, we run an experiment. Whichever model has the highest relative likelihood for fitting the experimental data is considered the winning model. Obviously, to find a winner, it is crucial to avoid ties in model fitting. To do so---before running the experiment---we look at the distributions of likelihoods that each models assign to every possible game dataset. Intuitively, we can expect much of the datasets to be associated with similar likelihoods from the three models. However, there is important information encoded in where the models' predictions diverge. Our procedure locates the experimental design(s) that accentuate the different model predictions, and this is represented by the experimental parameterization that maximizes the KL divergence between each model's likelihoods of observing particular datasets. We would then run that experiment and use the data collected to determine which model most likely describes the behavior we observe. Crucially, had we not run the optimal design, our ability to distinguish between---and eventually rank---the competing models might have been obscured. On the other hand, an experiment that was optimized for information gain may allow us to rule out suboptimal models of behavior after collecting only a small batch of data. This ability (in addition to the fact that the experiment will likely be cheaper and require fewer participants) gives ``information gain'' an important role in understanding and optimizing experimental design.

In the following section, we will detail the specifics of the experiment we are attempting to optimize. After this introduction to the experiment, the following two sections lay out the main contributions of the current paper. The first contribution is methodological; we describe two improvements to the optimal design procedure above. Then, we describe our method for comparing the outputs from our algorithm to experimental designs recommended to us from experts in behavioral economics and digital experimentation. The final section of this paper reports results from actually running several versions of this experiment in order to compare the predictions of experts to the output from our optimal design procedure.

\section{Sample Application: Imperfect Information Game and Behavioral Models (Types)}

Every study can benefit from an optimal experimental design. In the current work, we have chosen to optimize a two-player incomplete information game played over the course of three rounds, called the Stop-Go game \cite{El-Gamal1996EconomicalDesign}. This experiment is well-suited for an optimal design procedure because there are just two experimental parameters that define the gameplay, in addition to several model parameters. This is enough to create a vast number of possible outcomes, which makes an optimal experiment all the more necessary to run in order to learn from the data collected. Furthermore, we can directly compare our results against El-Gamal \& Palfrey (1996), taking into account potential differences in lab vs.~online behavior in the game.

The economic intuition for the Stop-Go game played over rounds is about learning and signaling of a given behavioral type within the population. In general terms, the assumption of homogeneity makes the game resemble the process of stereotype creation or, if played within a larger population over many rounds, norm formation. A more concrete economic application is the following: The game represents a number of situations in which the sender tries to manipulate the receiver with the ``Go'' signal but the receiver tries to out-think the sender's manipulation. For example, a business (the sender) is choosing whether or not to advertise (go or stop) and a customer (the receiver) is trying to decide which product to buy. The business wants the customer to get the worse deal while the customer wants the better deal. The business knows whether or not the product is a good deal and can choose whether or not to advertise. The customer sees only an advertisement but doesn't know whether the deal is actually good.

In the next subsections, we provide a brief explanation of the Stop-Go game and the behavioral models we would like to test on it. While an understanding of these details is important for this particular study, our optimal experimental design approach is fundamentally agnostic to both the specific experiment and the behavioral models it is applied to.

\subsection{Game Rules and Parameters to Optimize}\label{sec:stopgo_rules}

The Stop-Go game starts with nature randomly selecting the state of the world: State \textit{a} with probability $\pi$ and state \textit{b} with probability $1-\pi$ (these two probabilities are common knowledge). Player 1 (color ``Red'') is informed about the state of the world, while Player 2 (color ``Blue'') is not. Player 1 then chooses \textit{Stop} or \textit{Go}. \textit{Stop} ensures that both players receive a payout of 1, regardless of the state of the world.\footnote{ Following the design by El Gamal \& Palfrey (1996), Player 2 makes a decision even if Player 1 chooses \textit{Stop}.} If \textit{Go} is selected, it is Player 2’s turn to select \textit{Left} or \textit{Right}. Finally, Player 1 and Player 2 are given payouts that depend on the state of the world and on one of the design parameters to optimize (A). Fig.~\ref{fig:stopgo} shows the decisions tree for one round of a two-player Stop-Go game containing the two design parameters (A and $\pi$). Namely,

\begin{itemize}
    \item A is the maximum payout that either player can receive,
    \item $\pi$ is the probability that the state of the world is \textit{a}.
\end{itemize}

\begin{figure}[t]
\begin{center}
  \begin{tikzpicture}[scale=1.5,font=\footnotesize]
    \tikzstyle{level 1}=[level distance=15mm,sibling distance=35mm]
    \tikzstyle{level 2}=[level distance=15mm,sibling distance=12mm]
    \node(0)[solid node,label=above:{Nature}]{}
    child{node(1)[solid node,label=above:{1}]{}
      child{node[hollow node,label=below:{$(1,1)$}]{} edge from parent node[left]{Stop}}
      child{node[solid node,label=right:{}]{}
        child{node[hollow node,label=below:{$(0,2)$}]{} edge from parent
          node[left]{Left}
        }
        child{node[hollow node,label=below:{$(A,0)$}]{} edge from parent
          node[right]{Right}          
        }
        edge from parent node[right]{Go}}
      edge from parent node[left,xshift=-3]{$\pi$}
    }
    child{node(2)[solid node,label=above:{1}]{}
      child{node[solid node,label=left:{}]{} 
      child{node[hollow node,label=below:{$(2,0)$}]{} edge from parent
          node[left]{Left}}
        child{node[hollow node,label=below:{$(0,A)$}]{} edge from parent
          node[right]{Right}
        }
      edge from parent node[left]{Go}}
      child{node[hollow node,label=below:{$(1,1)$}]{}        
        edge from parent node[right]{Stop}}
      edge from parent node[right,xshift=3]{$1-\pi$}
    };
    \draw[dashed,rounded corners=10]($(1) + (.2,-1.23)$) rectangle ($(2) +(-.2,-1.8)$);
    \node at (0,-3) {2};
  \end{tikzpicture}
\end{center}
\caption{\textbf{Extensive-form representation of the game.} If Player 1 chooses \textit{Go}, Player 2 must make a choice between \textit{Left} and \textit{Right}. Possible payoffs of the game are $(0, 1, 2, A)$ with $A > 2$. The parameters to optimize are $A$ and $\pi$.}
\label{fig:stopgo}
\end{figure}

In El Gamal \& Palfrey (1996)'s experiment, A was set to 3.33 and $\pi$ to 0.5, generating payoffs as shown in Table \ref{fig:payouts}.

\begin{table}[H]
\centering
\begin{tabular}{|| c | c | c | c ||} 
 \hline 
 \multicolumn{4}{||c||}{\textbf{Payoff Tables} when A = 3.33 (\$10.00)} \\
 \hline\hline
 \multicolumn{2}{|c|}{\textbf{Payoff Table World \textit{a} ($\pi$)}} &
 \multicolumn{2}{|c|}{\textbf{Payoff Table World \textit{b} ($1-\pi$)}} \\
 \hline\hline
 \textbf{Left} & \textbf{Right} & \textbf{Left} & \textbf{Right}  \\ 
 \hline
 Player 1: \$0.00 & Player 1: \$10.00 & Player 1: \$6.00 & Player 1: \$0.00 \\ 
 Player 2: \$6.00 & Player 2: \$0.00 & Player 2: \$0.00 & Player 2: \$10.00 \\ 
\hline
\end{tabular}
\caption{\textbf{Payoff table of the game.} Assuming Player 1 chose Go, the payoffs for both players depend on the state of the world (\textit{a} or \textit{b}), and on the choice of Player 2 (Left or Right).}
\label{fig:payouts}
\end{table}

The game is played in a group of 10 players for three rounds. Each player is assigned a role (1 or 2) which is kept fixed throughout the whole game. Each round, players are rematched so that they meet a new partner playing the other role (``perfect stranger'').

\subsection{Models of Behavior and Parameters to Optimize}

The original paper by El-Gamal \& Palfrey (1996) introduced the parameter $\pi_{per}$ for the participants' perception of the actual probability $\pi$ and tested the following three models of behavior:\footnote{ Wording as used in El-Gamal \& Palfrey (1996). See Appendix for a more detailed description.}

\begin{enumerate}
    \item \textit{Model 1}: Each individual \textit{i} plays the Bayes-Nash Equilibrium of the game, defined by ($\pi_{per}, A, \epsilon$).
    \item \textit{Model 2}: Player 2 does not update $\pi_{per}$ following \textit{Go}, and this is common knowledge.
    \item \textit{Model 3}: Individuals use fictitious play to construct beliefs about opponents' play.
\end{enumerate}

Each of the three models takes as inputs the two experimental parameters---$A$ and $\pi$---described in the previous section and three model parameters: $\alpha$ (learning speed), $\epsilon$ (tremble rate), and $\delta$ (accuracy in perception of probability $\pi$---it generates $\pi_{per}=\pi\pm\delta$). Table \ref{table:params} provides a summary of both experimental and model parameters, their meanings, values, and ranges. The goal of our Bayesian optimal experimental design is now to find the point in the experimental parameter space ($A$ times $\pi$) where the likelihoods of three models differ the most.

\begin{table}[H]
\centering
 \begin{tabular}{|| c | c | c | c | c ||} 
 \hline 
 \textbf{} & \textbf{Definition} & \textbf{Parameter Type} & \textbf{Range} & \textbf{Discrete Values} \\
 \hline\hline
 $A$ & maximum payout to participants & experimental design & 2.00--6.00 & 20 \\
 \hline
 $\pi$ & probability of state of the world, \textit{a} & experimental design & 0.1--0.9 & 20 \\
 \hline
 $\epsilon$ & tremble rate & model & 0.00--1.00 & 34 \\
 \hline
 $\alpha$ & learning rate & model  & 0.00--1.00 & 34 \\
 \hline
 $\delta$ & misperception of $\pi$ & model & 0.00--0.20 & 7 \\
 \hline
 \end{tabular}
 \caption{\textbf{Parameters in the game.} 
   In order to define the \textit{optimal} experiment, we need to assign likelihoods to the observations that are generated under each possible combination of these parameters. In our computations, continuous parameters are approximated using the number of discrete values shown in the last column.}
 \label{table:params}
\end{table}

\section{Optimal Experimental Design: Implementation}\label{sec_oed_implementation}

In order to determine the optimal experimental design that maximizes information gain, we need to compute how likely it is to observe different outcomes. The first step in this calculation is to enumerate \textit{every possible dataset} we might observe when we run our experiment. We define the behavioral responses observed at the smallest unit of analysis in the game as a single \textit{outcome}. In our case, this is a pair of players playing one round of the game. Following Fig.~\ref{fig:stopgo}, this gives us eight possible outcomes: \textit{a}-go-left, \textit{a}-go-right, \textit{a}-stop-left, \textit{a}-stop-right, \textit{b}-go-left, \textit{b}-go-right, \textit{b}-stop-left, and \textit{b}-stop-right.\footnote{ We closely follow the implementation of the Stop-Go game by El Gamal \& Palfrey (1996), where Player 2 makes a decision even if Player 1 chooses Stop.} The ensemble of all outcomes for all players for all rounds is a \textit{dataset} (i.e., the data obtained after running the experiment once). The total number of possible datasets is equal to the number of permutations of outcomes with replacement. Even in a basic imperfect information game like ours, the number of possible datasets quickly becomes very large. For example, with only one round and five pairs of players, that number is 32,768. For three rounds, the total number of unique datasets is equal to $32,768^3\approx3.5^{13}$. This is a great deal of likelihoods to compute! However, in this count, some datasets are ``duplicates'' in the sense that they contain the same distribution of outcomes but the players generating those outcomes switched positions. Depending on how the models actually assign likelihoods to outcomes (e.g., if and how they take into account the history of past outcomes, as would learning models) and on the rules for matching players in the game, the total number of likelihoods to compute can be greatly reduced. However, in our specific case, this is not possible, as games with strategic interactions, information asymmetries, or network effects require us to keep track of the sequence of outcomes experienced by each participant. 

\begin{figure}[t]
\centering
\begin{subfigure}{.326\textwidth}
  \centering
  \includegraphics[width=1.0\columnwidth]{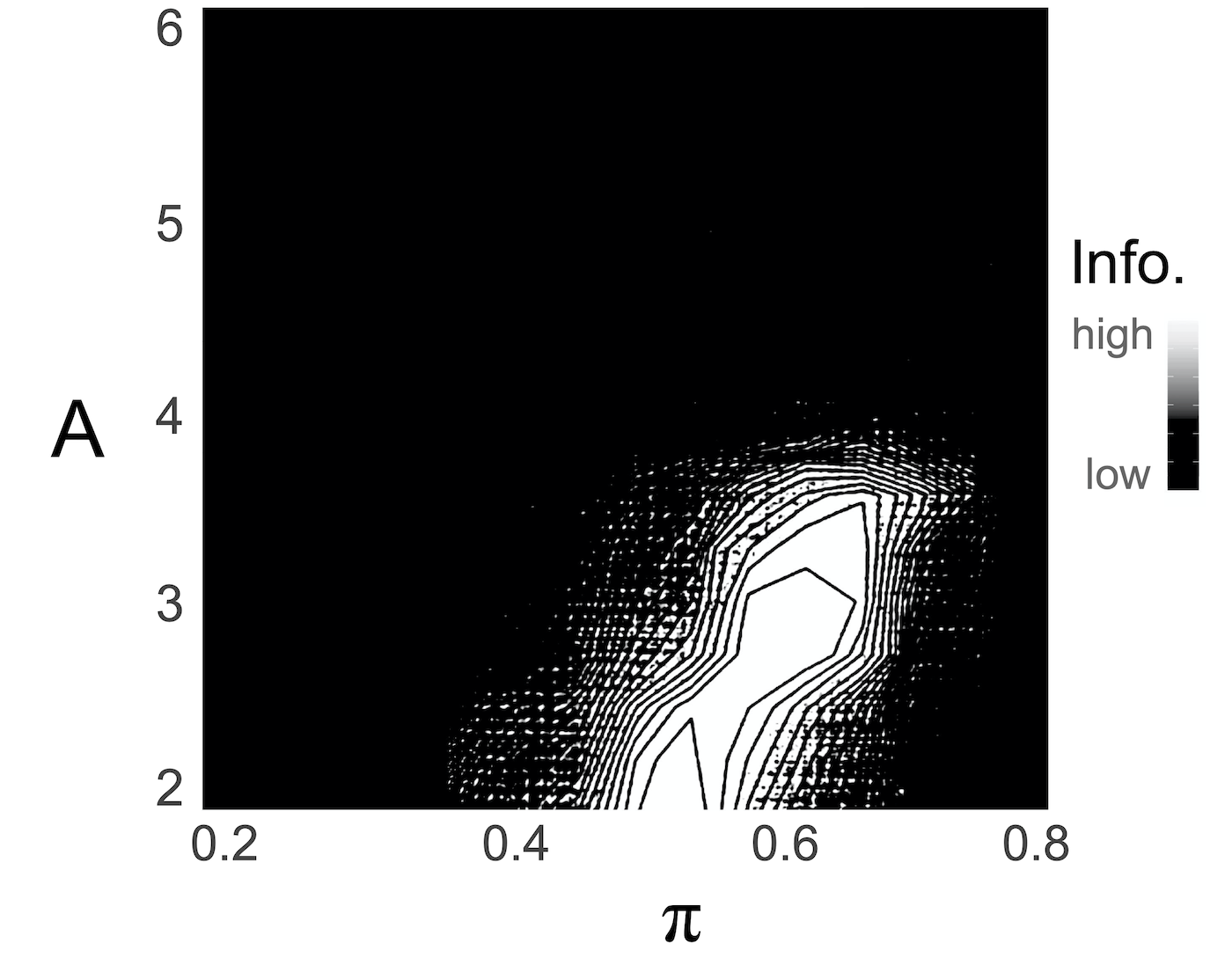}
  \caption{Original method, El-Gamal \& Palfrey (1996)}
  \label{fig:orig}
\end{subfigure}
\begin{subfigure}{.326\textwidth}
  \centering
  \includegraphics[width=1.0\columnwidth]{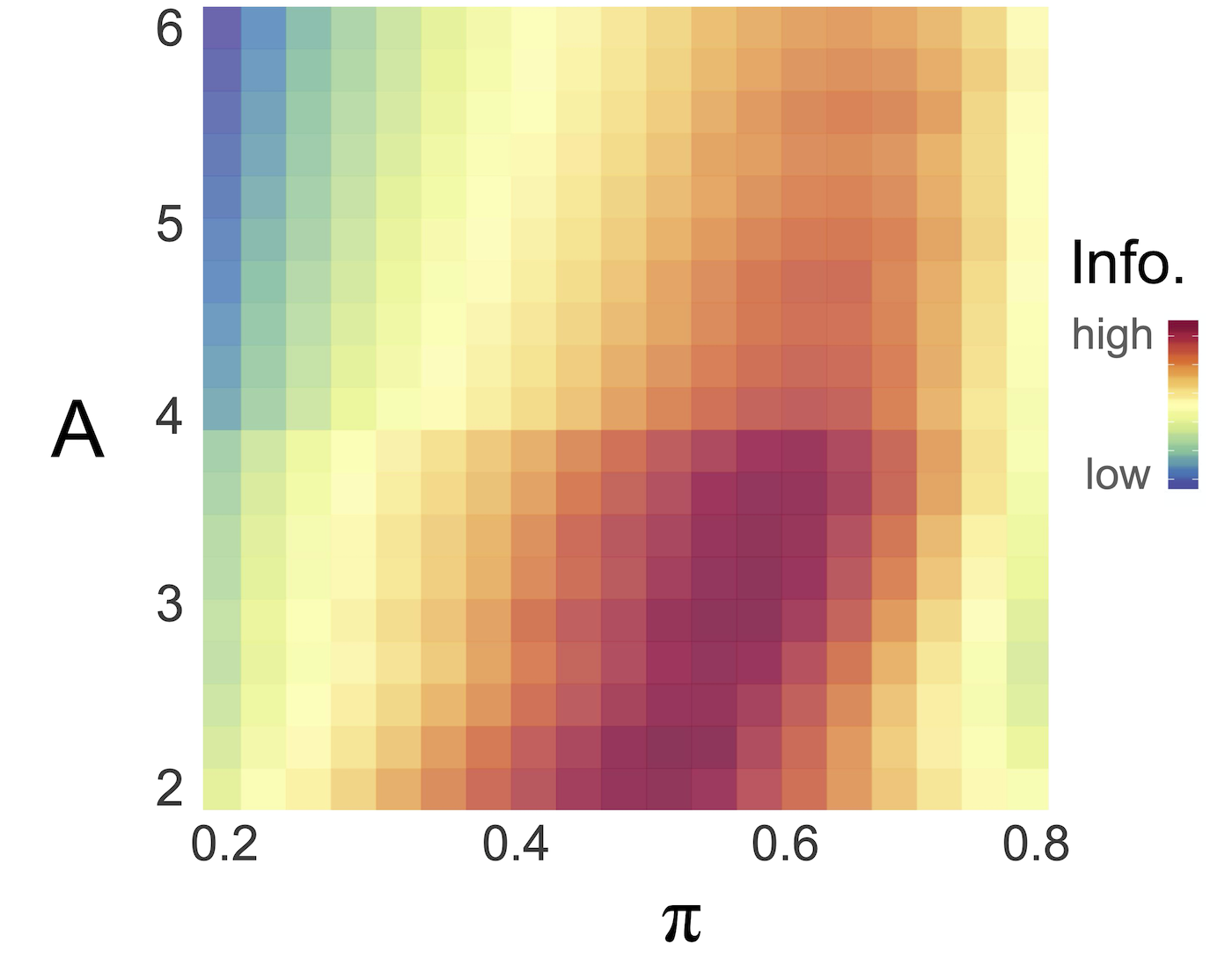}
  \caption{Replication of method from El-Gamal \& Palfrey (1996)}
  \label{fig:replic}
\end{subfigure}
\begin{subfigure}{.326\textwidth}
  \centering
  \includegraphics[width=1.0\columnwidth]{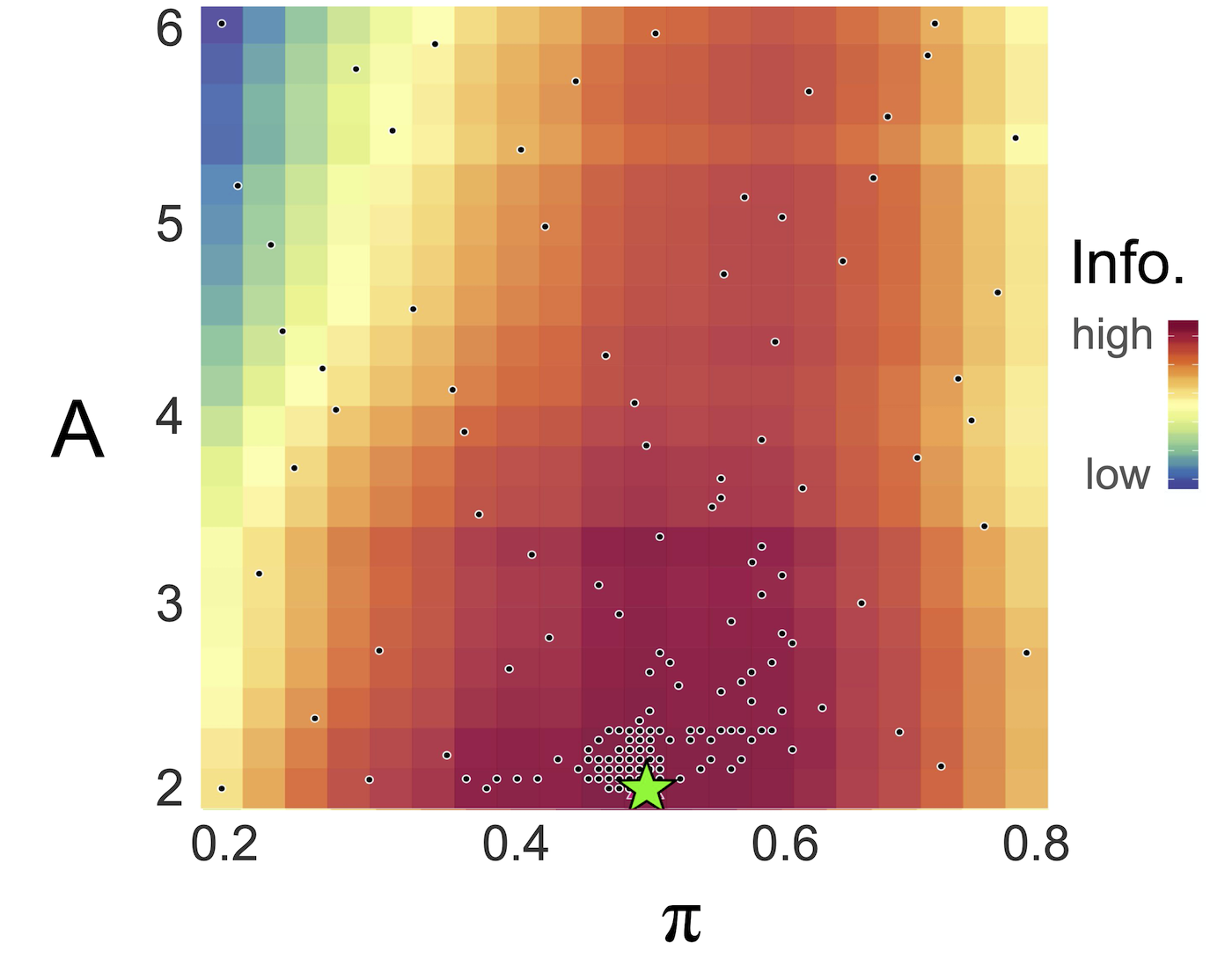}
  \caption{Improved search method using Parameter-Sampled GPUCB-PE}
  \label{fig:gpucbpe}
\end{subfigure}
\caption{\textbf{Information surfaces for experimental design.} (a) Information surface originally presented in El-Gamal \& Palfrey (1996) (b) Information surface generated by replicating the optimal design procedure in El-Gamal \& Palfrey (1996); (c) Information surface generated using Parameter-Sampled GPUCB-PE, where the points shown represent coordinates searched by the algorithm, and the green star representing the point corresponding to the experiment that is predicted to produce the maximum information gain. Note: as described in El-Gamal \& Palfrey (1996), the experiment with the maximum information is expected to occur when $A$ reaches its minimum value, which approaches (but is not equivalent to) $A=2.0$.}
\label{fig:surface_compare}
\end{figure}

Recall that this optimal design procedure is meant to distinguish competing models of behavior, and in order to determine the optimal experiment, we must first use each competing model to assign a likelihood to each unique dataset, \textit{for each combination of experimental design and model parameters}. In the current study, we use 3 model parameters ($\epsilon$, $\alpha$, $\delta$). Each of these parameters needs to be integrated out when calculating the likelihood of an individual dataset. We discretized them to test 34 different values of $\epsilon$, 34 different values of $\alpha$, and 7 different values of $\delta$ (see Table \ref{table:params}). Each value of $\delta$ leads on average to 7 values of the model parameter, $\pi_{per}$, hence $34 \times 34 \times 7 \times 7=56,644$. This generates 56,644 unique combinations of \textit{model} parameters for each \textit{experimental} parameter. There are 2 experimental design parameters ($A$, $\pi$). In the default, brute-force approach, we evaluated them in a grid of $20 \times 20=400$ experimental parameter combinations. The output of an optimal design procedure such as this one is an \textit{information surface} that represents the multi-dimensional (in our case, two-dimensional) grid of every possible experiment and its associated information gain from taking the KL divergence between the dataset likelihoods of the competing models (Fig.~\ref{fig:surface_compare}). In our case, the total number of computations necessary to determine the optimal design of our experiment is equal to $400 \times 56,644 \times 3.5^{13}\approx8^{20}$.

\subsection{Algorithmic Improvements for Optimal Design}\label{sec:alg_improvements}

Running the total number of likelihood calculations proved computationally infeasible on a super-computing cluster, even for the coarse grid of experimental design parameters we chose.\footnote{ In fact, we could construct the information surface for choosing the optimal design parameters only for a simpler two-player version of our game, and it took approximately 72 hours on the following super-computing cluster: Four hundreds parallel R v.3.x jobs, distributed across 56-core x86 64 Little Endian Intel(R) Xeon(R) cpus (E5-2680 v4 @ 2.40GHz; L1d cache: 32K, L1i cache: 32K, L2 cache: 256K, L3 cache: 35840K).} To address this, we introduce two improvements to the Bayesian optimal design procedure.\footnote{ All code is available at \url{http://github.com/shakty/optimal-design}.} 

\begin{enumerate}
    \item \textit{Which} experimental designs are evaluated? We use Gaussian Process search to adaptively select the next coordinate in the information surface to evaluate (Section \ref{sec:gpucbpe}).
    \item \textit{How} are points evaluated? We sample datasets (by uniformly sampling \textit{model} parameters and simulating datasets) instead of spanning through all possible datasets (Section \ref{sec:param_samp}).
\end{enumerate}

The first improvement gets around the requirement to grid-search every combination of experimental design parameters in order to construct the information surface. The second improvement gets around the costly step of calculating the likelihoods of all possible datasets under a given model, which was previously needed for computing the information associated with each combination of experimental design parameters.

\subsubsection{Gaussian Process Upper Confidence Bound-Pure Exploration}\label{sec:gpucbpe}

Given the computationally taxing nature of this systematic grid-search technique for finding the optimal experimental design, we implemented an adaptive search technique used in artificial intelligence research to speed up the construction of the information surface. This approach leverages recent algorithmic developments for adaptive search from machine learning, known as Gaussian Process (GP) regression \cite{Rasmussen2005, gortler2019a}. The basic premise behind GP is that there are often local correlations among data points, and having prior knowledge about a given data point can inform your posterior belief about ``nearby'' data. When dealing with multidimensional data, this simple assumption proves to be quite useful for generating insights about unobserved data, allowing for the automated generation of predictions about the entire search space without needing a systematic grid search.

The search algorithm we use is known as Gaussian Process Upper Confidence Bound Pure Exploration (GPUCB-PE), and it is a way of reconstructing the contours of a landscape in an adaptive manner, while also minimizing \textit{regret}. Here, regret is defined as the difference between the maximum observed value until a given timestep and the final maximum observed value \cite{Stefanakis2014CanApproach, Contal2013}. GPUCB-PE was introduced as a variant of GPUCB \cite{Contal2013} and was designed to encourage more exploratory searches (hence the inclusion of ``Pure Exploration'')---behavior that is especially effective when the landscape being searched is especially large or otherwise complex. 

Recall that our search task is on an information landscape where each point is a possible combination of experimental design parameters, each of which are associated with an information value (how much information we would gain from running an experiment with those design parameters). Intuitively, one can imagine a subset of experimental designs that are simply \textit{uninformative} (i.e., they are likely to generate datasets that will be unable to distinguish between our competing models), and it would be computationally wasteful to exhaustively search the experimental design landscape in this region. Leveraging this allows the algorithm to focus its search on parameter combinations that are likely to be good, while reducing the evaluation of parameter combinations that are likely to be bad. However, this poses an explore-exploit dilemma: how can the algorithm ensure that there are not hidden pockets of high-value parameter combinations in otherwise low-value regions? Herein lies the power of the GPUCB-PE algorithm in particular: the \textit{pure exploration} (PE). This variant of Gaussian process search explicitly balances the exploitation of high-value regions of the landscape while periodically searching in regions with high \textit{uncertainty} (see SI \ref{appendix:GPUCBPE} for more detail), thereby minimizing the chances of overlooking a high-value point hidden in a low-value region of the landscape.

In summary, this first algorithmic improvement allows us to quickly uncover the experimental parameter combination with the highest information value while generating inferences about the information value of the points it has not yet searched \textit{and} avoiding unnecessary searches in low information value regions. It does this by:
\begin{enumerate}
    \item Initializing a multi-dimensional surface of coordinates that represent possible combinations of experimental parameters (in this case, combinations of $A$ and $\pi$ values).
    \item Using Equation \ref{eq:DKL}, calculate the KL divergence between the model likelihoods of the \textit{n} competing models for a sample of initial points. These initial points can be IID random points in the landscape, but we chose to distribute the initial points using a quasi-random spacing technique known as Sobol sequencing \cite{Sobol1998} which achieves better initial coverage.
    \item Proceeding with the GPUCB-PE algorithm, iteratively selecting points in the landscape and calculating the information gain at each of those points, while also using Gaussian Process regression to \textit{infer} the expected information gain and confidence bounds of every point that has not yet been searched. This algorithm alternates between selecting (i) the point with the highest expected information gain \textit{plus} its upper confidence bound (the ``UCB'') and ii) the point---within a region of eligible points---with the largest confidence bounds (i.e., the largest uncertainty, detailed in Appendix \ref{appendix:GPUCBPE} and \cite{Stefanakis2014CanApproach}). The two types of points that this algorithm selects address the \textit{exploitation} and \textit{exploration} of the landscape, respectively. 
    \item After each new point has been searched, determining if the landscape satisfies the \textit{stopping rule}, which is based on the similarity between the current reconstruction of the landscape and the previous reconstruction. If the two landscapes are repeatedly more than 99.9\% similar to each other according to a modified Spearman rank correlation, the algorithm stops and the final information surface is output.
\end{enumerate}

This method enables us to reconstruct an accurate information surface (1) without needing to span the grid of all possible experiments, (2) with more precision than a grid-searched landscape, as maxima may be found inside the tiles described by the grid (see Section \ref{appendix:GPUCBPE} in the Appendix), and  (3) all while minimizing the \textit{regret} of the search process. For the interested reader, we recommend foundational work on the regret of GPs, showing that GPs will converge to the optimal solution at an approximate rate of $\mathcal{O} \Big(\frac{1}{\sqrt{t}} \Big)$ \cite{srinivas2009gaussian}, as well as more recent work that accommodates the dimensionality, $d$, of the search space, showing that GPs converge to the optimal solution at a rate of $\mathcal{O}\Big(e^{\frac{-\tau t}{(\ln t)^{d/4}}}\Big)$ \cite{de2012exponential}. While simply being regret-minimizing does not negate the challenge that a complex search space might play a role in ultimately converging on the optimal experimental design, it illustrates the flexibility and power that GPs have in addressing this problem.

In the next section, we introduce the second algorithmic improvement used in this work and define our novel search algorithm: the Parameter-Sampled GPUCB-PE.

\subsubsection{Sampling Parameters to Sample Datasets}\label{sec:param_samp}

While GPUCB-PE addressed \textit{which} points in the information landscape need to be evaluated, our second improvement addresses \textit{how} the information value of each point is evaluated; namely, we introduce a model parameter sampling procedure that allows us to avoid the costly calculation of model likelihoods for every possible dataset. Recall that there are two main classes of parameters in this work: \textit{experimental design} parameters and \textit{model} parameters (Table \ref{table:params}). We use GPUCB-PE to find the most informative combination of experimental design parameters. In order to arrive at an information value for each point in the experimental design landscape, we compute a KL divergence between the likelihoods that each model assigns to every possible dataset, for a specific experimental design. Assigning a likelihood to \textit{every} possible dataset is immensely costly, and as the number of participants and/or the number of rounds in an experiment increases, this computational complexity becomes prohibitively large.

We introduce a simple and effective solution for this problem, which we will refer to as the \textit{Parameter-Sampled GPUCB-PE}. The basic algorithmic procedure is identical to the GPUCB-PE approach described in Section \ref{sec:gpucbpe} (i.e., iteratively searching combinations of experimental parameters, honing in on those with high information values until convergence). However, when it comes to the calculation of the information value---taking the KL divergence between each model's likelihoods of the approximately $3.5^{13}$ datasets---we show that we can achieve the same performance by comparing the likelihoods of only a fraction of all possible datasets in the KL divergence calculation. 

Parameter-Sampled GPUCB-PE lets us avoid spanning all $N$ possibly-observed datasets by uniformly sampling $n_{s} \ll N$ combinations of model parameters and using them to generate $n_s$ synthetic datasets. In essence, each of the $n_s$ samples of model parameter combinations simulates the datasets that \textit{would} be generated \textit{if} participants were making decisions under a given model with the selected model parameters. In full, this model parameter sampling procedure entails the following steps:

\begin{enumerate}
    \item For a given model, draw a parameter value uniformly at random for each model parameter.\footnote{ Non-uniform sampling can be used when the experimenter has prior over the distribution of model parameters.} As an example, consider Model 1, which includes three parameters: $\epsilon, \alpha, \delta$. This step simply draws a random value for each parameter ($\epsilon_i, \alpha_i, \delta_i$), from each of their respective ranges (see Table \ref{table:params} for the ranges of values that these parameters can take).
    \item For each model being compared, sample a dataset that might be observed if the simulated participants behaved according to the model parameters $\epsilon_i, \alpha_i, \delta_i$, for all rounds. This is done by inserting the sampled parameters inside each of the model likelihood functions (Section \ref{appendix:models} in the Appendix) and sampling a dataset from the resulting distribution of datasets.
    \item Repeat Steps 1 and 2 for a sufficiently high number of times, $n_s$. Each model being compared will ultimately have its own sample of $n_s$ datasets.
    \item For each unique dataset, compute the fraction of times it shows up in each of the models' samples, thereby creating a histogram of dataset frequencies. These histograms are meant to approximate the true distribution of likelihoods of datasets and, as such, we use them to calculate the KL divergence.
\end{enumerate}

We sample $n_s=10,000$ datasets per search---several orders of magnitude less than what was required before. Not only does this approach reduce the number of computations by several orders of magnitude, we also found that sampling around $1,000$ datasets was sufficient to 
(1) qualitatively reconstruct the information landscape, 
(2) make the algorithm satisfy the stopping rule, and 
(3) most importantly, find the coordinate of the optimal design on the information surface (Appendix \ref{sec:ps_gpucbpe_regret} for further analysis). 

It should be noted that the Parameter-Sampled GPUCB-PE has no prior knowledge of neither the underlying models nor their behavioral assumptions. Even so, it is able to reliably reconstruct information surfaces, proving itself to be a very versatile approach to create informative experiments in any knowledge domain. In addition, while we use this Parameter-Sampled GPUCB-PE approach in order to find the maximally informative experimental design, this algorithm can be used to create an analogous ``information surface'' corresponding to any measure that we may want to optimize (e.g., experiment cost, treatment effect).

Together, the two improvements address and improve two key bottlenecks in the Bayesian optimal experimental design procedure: \textit{which} points get evaluated when searching for the most informative experiment and \textit{how} the information value of each point gets assigned. 

\begin{figure}[t!]
\centering
\includegraphics[width=1.0\columnwidth]{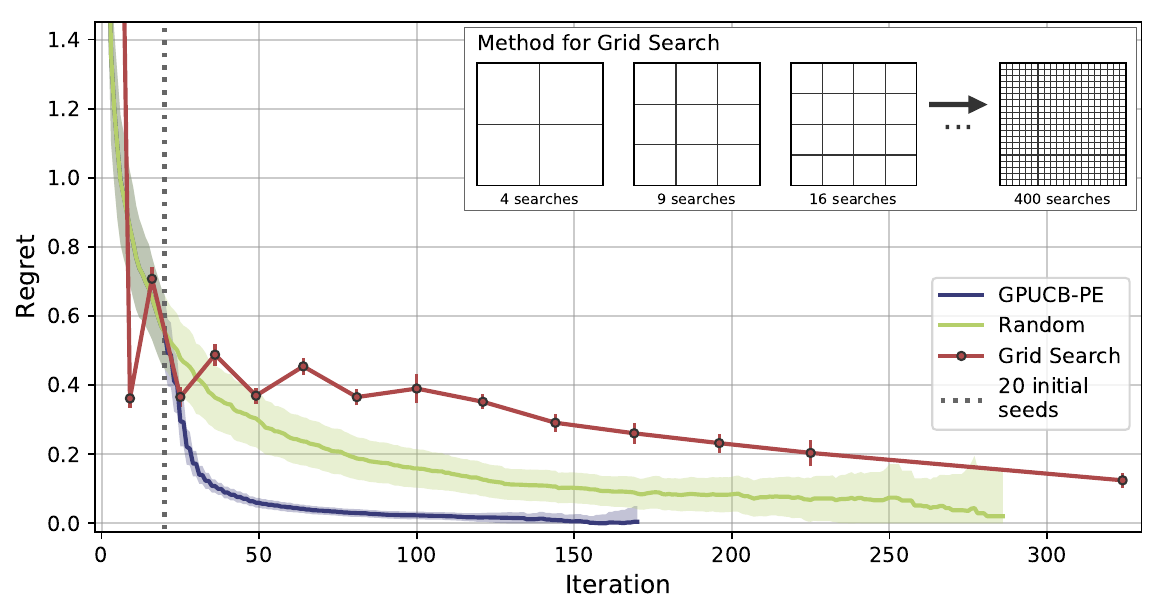}
\caption{\textbf{Comparison of algorithm performance measured by regret.} Here we compare the performance of three different search methods for finding the optimal experiment: Random search, Grid search, and GPUCB-PE. We plot the mean regret over time for each method (averaged over 200 runs), where regret is defined as the difference between the maximum value uncovered by the algorithm and the average global maximum. We show the 95\% confidence interval with translucent bands around the curves (error bars for Grid Search). For each search method, we mimic the steps of our Parameter-Sampled GPUCB-PE algorithm. That is, we sample datasets instead of exhaustively searching the parameter space: for every \textit{experimental} parameter that the algorithm selects (i.e., for every coordinate searched on the information surface), 10,000 \textit{model} parameters are sampled, and a synthetic dataset is generated using these sampled experimental and model parameters, for each of the three models being compared. These sampled datasets are then used to assign an information value to each coordinate in the information surface. The Parameter-Sampled GPUCB-PE algorithm finds a solution close to the global maximum almost immediately, whereas Random and Grid Search require more searches in order to find a value close to the global maximum.}
\label{fig:algo_performance}
\end{figure}

\subsection{Performance Evaluation}\label{sec:performance_eval}

We compared the effectiveness of our Parameter-Sampled GPUCB-PE algorithm to a grid search of the landscape and a random search (Fig.~\ref{fig:algo_performance}). The algorithmic improvements we have made to the optimal design procedure laid out in El-Gamal \& Palfrey (1996) constructs an information surface that maintains the same general contours and precisely the same global maximum value as the original algorithm (compare Fig.~\ref{fig:gpucbpe} and \ref{fig:replic}). This allows us to rapidly understand the role that particular combinations of experimental design parameters play in determining how \textit{informative} an experiment is. Furthermore, it is able to do it considerably faster than the traditional grid-search approach, evaluated here by iteratively constructing larger size grids (see inset in Fig.~\ref{fig:algo_performance}). A key limitation of the grid-search approach, in addition to being slow, is that the coordinate of the global maximum value could--and indeed does--fall between the cells of the grid. Thus, even with a very fine grid, it never finds a solution as good as the GPUCB-PE. The GPUCB-PE algorithm performs significantly better than Random and Grid search, achieving the lowest regret and achieving it fast. Statistical analysis confirms that the regret of GPUCB-PE is significantly lower than Random starting iteration 30 until Random search incidentally discovers the true global maximum at approximately iteration 275 (t-test, comparison of means $p<10^{-10}$ at iteration 30).

\section{Wisdom of Experts}\label{sec:wisdom}

There are a number of ways to assess the \textit{effectiveness} of the optimal experimental design protocol we have laid out. As we have previously described, our algorithm finds the Bayesian optimal experiment that maximally distinguishes competing models of behavior. However, a persistent question that arises in this approach is whether or not the output from this optimal design procedure would even be useful to experimentalists in academia or industry. Is the algorithm simply generating outputs that would be obvious to a trained experimenter? We sought out experts in behavioral economics and digital experimentation to answer this question, and we asked them to complete a survey where they provided us an estimate of what experiment \textit{they} would run if they were in control of the design.

We first described the task: these experts were to suggest parameters for an experimental design that would maximally distinguish the likelihoods of four different models of behavior. We then collected demographic information about the respondents (their gender, field of expertise, academic position, and country of employment). We then described the imperfect information game, including the same language and figures from Section 3 and Fig.~\ref{fig:stopgo} (for precise wording, see Appendix). After describing the four models the experiment was meant to distinguish, we asked the participants to suggest to us the precise values for $A$ and $\pi$ that would isolate and distinguish the different models from one another. These values were presented as text boxes that were constrained by the possible parameters for $A$ and $\pi$. The participants then reported their \textit{confidence} in each of those predictions (from 1 - not confident to 5 - very confident), as well as indicating which model they believed would eventually perform best after the experiment was ultimately run.

\subsection{Recruitment Strategy}
In order to assemble a group of experts who were reasonably-suited to perform this task, we sought out researchers who were familiar with experimental design as well as common behavioral models such as Nash Equilibrium. We preferred to look for researchers who were currently faculty in a department related to behavioral economics. These criteria led us to choosing to email researchers from the following areas:
\begin{itemize}
    \item Authors of work presented at the CODE@MIT, a conference on digital experimentation (2014, 2015, 2016, and 2017).
    \item Attendees of the ``Behavioral/Micro'' course at the National Bureau of Economic Research (NBER) Summer Institute (2012, 2013, 2014, 2015, 2016, and 2017).
    \item Authors of work presented at the Economic Science Association's North American annual conference (2015, 2016, and 2017).
    \item Boston-area faculty in economics (including Harvard University, Massachusetts Institute of Technology, Northeastern University, Boston University, Boston College, University of Massachusetts Amherst, and University of Massachusetts Boston).
\end{itemize}

This process generated 811 names, to whom we sent an initial email on 12/01/2017 (23 emails bounced back, resulting in a total of 788 successfully delivered emails). This initial email served to introduce the project, provide information about Northeastern University's IRB protocol, provide a link to the survey itself, and allow recipients to opt-out of further communication. After one week, a reminder email was sent to those who had not completed the survey. After another week, a thank-you email was sent to those participants who had completed the survey. This protocol was adapted from methods used in \cite{DellaVigna2016WhatForecasts}. In total, 147 participants began the survey (i.e., opened the browser window and agreed to the informed consent), and 55 participants actually finished the survey. There is no difference across respondents who started the survey and those who dropped out after beginning it. We find that more senior invitees are less likely to accept the survey invitation. However, the final sample is mainly composed of experienced researchers---those who are probably the most interested in the subject---and contains only 11 PhD students. Full details in Sec. \ref{sec_survey_dropouts} of the Appendix.

\subsection{Expert Prediction Results}
We report the major descriptive results here. For full survey results, see Table \ref{fig:experts} in the Appendix. The median response time was approximately 11 minutes. The mean response time was 105 minutes, although after removing participants who took over 6 hours to complete the survey (we assume that they were not actively working on the survey for that amount of time), this value became closer to 16 minutes. Of the 18\% of participants who began the survey, 39\% ended up finishing, making for a 7\% response rate overall.

\begin{figure}[t]
    \begin{subfigure}{.49\textwidth}
        \centering
        \includegraphics[width=0.99\columnwidth]{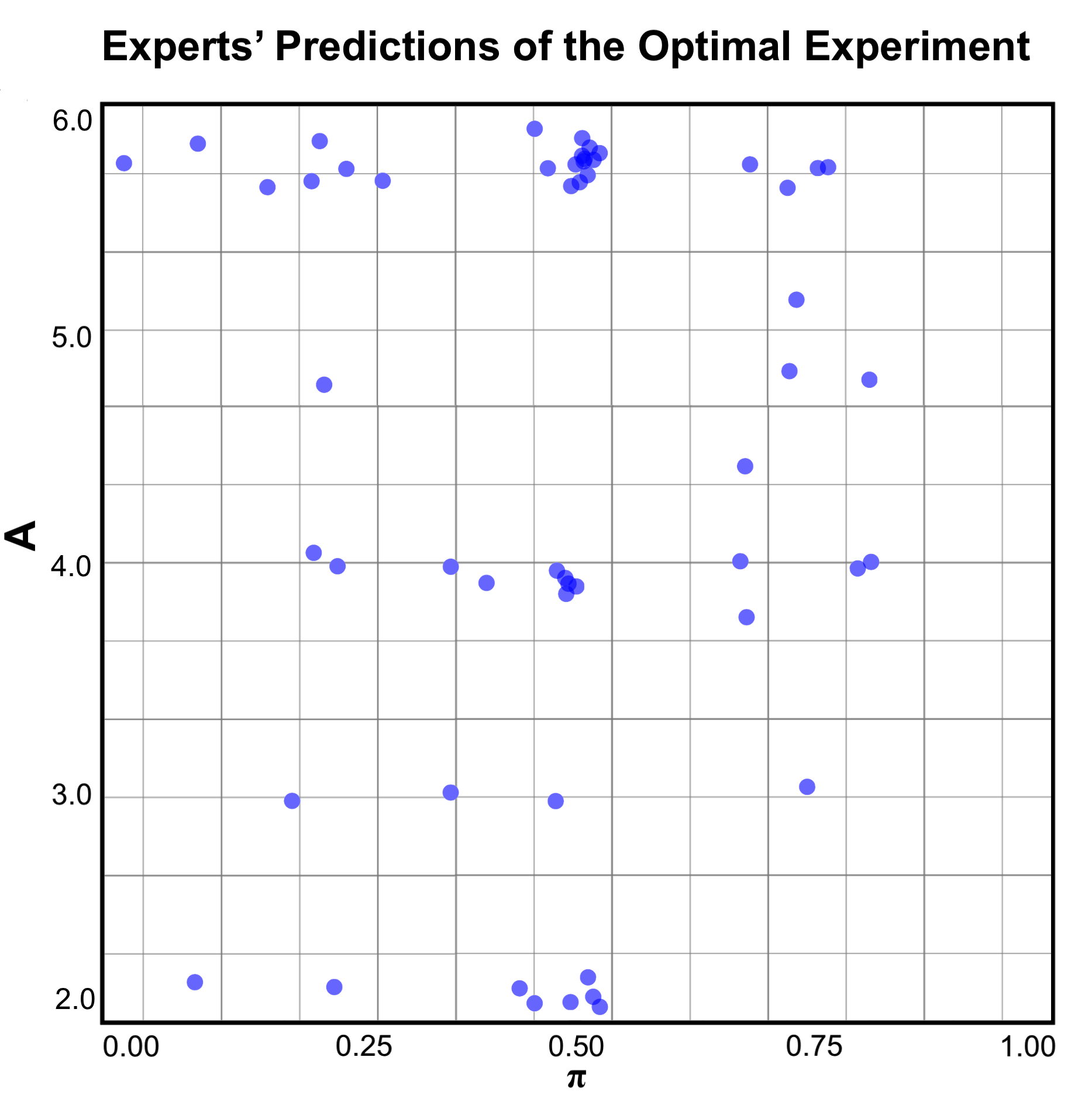}
        \caption{Experts' optimal experiment predictions (n=55).}
        \label{fig:experts_predictions}
    \end{subfigure}
    \begin{subfigure}{.49\textwidth}
        \centering
        \includegraphics[width=0.99\columnwidth]{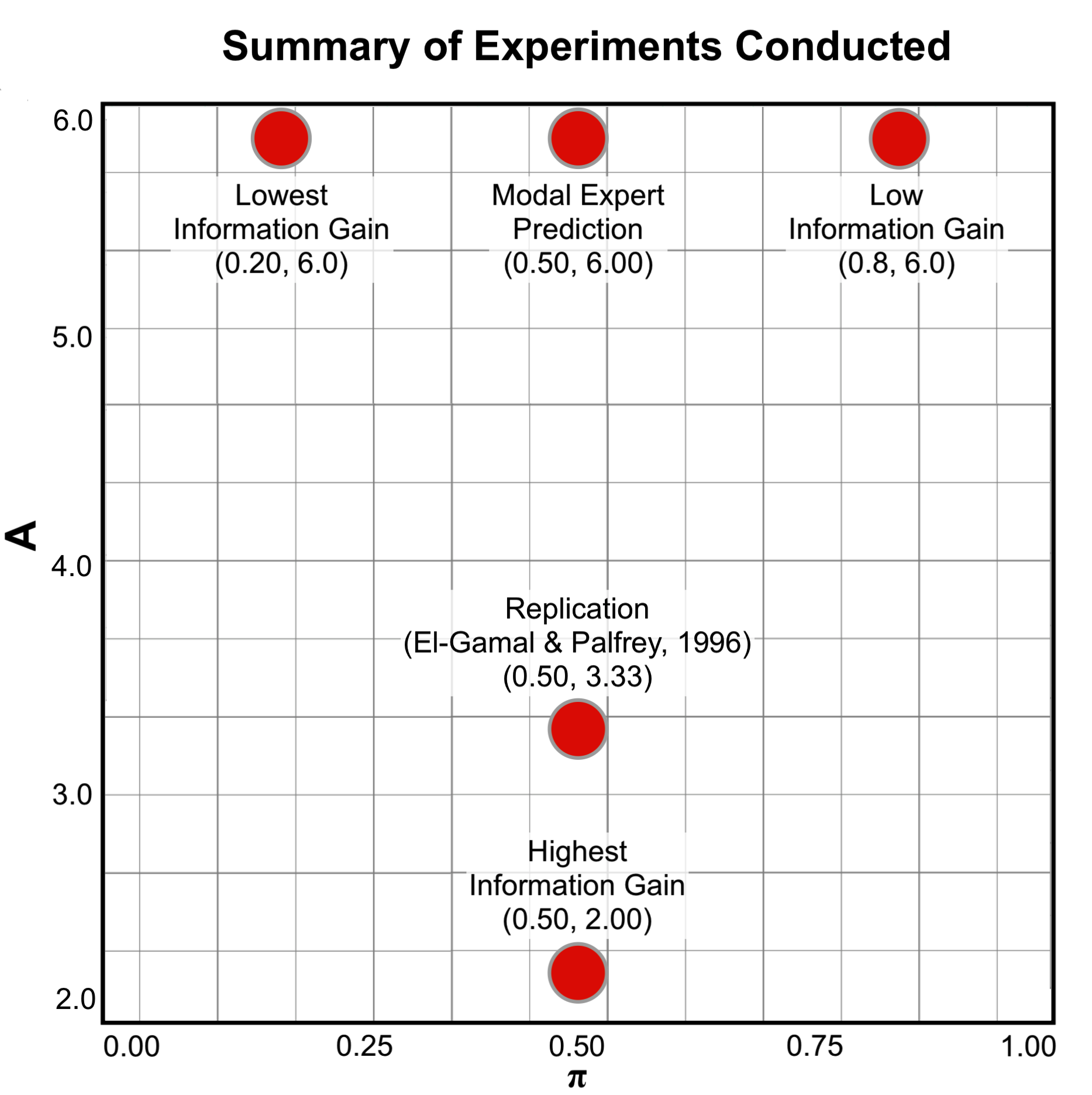}
        \caption{Final description of experiments run.}
        \label{fig:final_experiments}
    \end{subfigure}
    \caption{\textbf{Expert predictions and experiments chosen.} (a) Raw data from the experts survey showing both the modal prediction of $A = 6.0$ and $\pi=0.5$, but more importantly, the inherent noise and uncertainty that the respondents showed (data shown with small jitter to avoid over plotting); (b) These five points in the experimental design space were ultimately the ones that were tested on Amazon Mechanical Turk. Note: the experiment with highest information gain is when $\pi=0.5$ but \textit{approaches} $A=2.0$. We report results from the design $(\pi=0.5, A=2.0)$ as human participants would be unable to distinguish $A=2.000$ from $A=2.001$, for example.}
\label{fig:experimentsRun}
\end{figure}

This survey generated three main takeaways. First, and most importantly, there appeared to be little agreement among the various respondents regarding what was the optimal experiment to run, shown in Fig.~\ref{fig:experts_predictions}. This is validated by the low median self-reported confidence in their predictions about both $A$ and $\pi$ (a median of ``2 - somewhat not confident'' for both $A$ and $\pi$). 

Second, while the estimates spanned much of the experimental design space, the modal value was $\pi=0.5$ and $A=6.0$. This corresponds to a point on our information surface that would be expected to generate less information gain than the optimal design. The effect of running this suboptimal experiment might be that the competing models would appear to be less distinct, at which point we would need more data in order to distinguish the most likely model. 

Third, this modal value suggests an interesting mechanism for how experts choose experimental designs when they are uncertain. While the most commonly chosen value for $\pi$ (0.5) was the same as the optimal value, the experts surveyed were inclined to recommend a experiment that would pay higher incentives as the optimal one ($A=6.0$) making the compensation of research subjects \textit{more expensive} (if at least one game results in \textit{Go-Right}). That is, experts appeared to believe that giving research participants a higher reward in an experiment would help to clarify greater differences among the four competing models.\footnote{ Note: the optimal experiment we report in Figure 3---$A\approx2.0$ and $\pi\approx0.5$---is generated from an information surface comparing three models. The same coordinate is the optimal experiment when we include all four models. See Fig.~\ref{fig:four_model_surface} in the Appendix for the full information surface.} Indeed, it is generally accepted in experimental economics that greater rewards can increase participants' effort and attention during an experiment \cite{Knez1994, Hertwig2001}, even if some contrary evidence is also found \cite{Kachelmeier2005}. Additionally, experts may have made further assumptions about the role of different parameters of the behavioral models (e.g., participants' actual tremble rate). Nevertheless, according to the landscape generated by our theory of optimal design, lower values of the payoff parameter, $A$, tend to increase the informativeness of our experiment, not higher values.

We chose to include the modal value of $\pi=0.5$ and $A=6.0$ from these results as an experiment to test when we eventually tested this optimal design procedure.

\begin{figure}[t!]
    \centering
    \includegraphics[width=1.0\columnwidth]{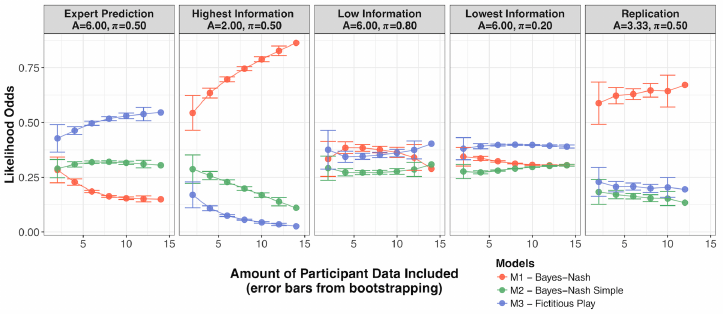}
    \caption{\textbf{Likelihood odd ratios of models according to real participants behavior.} The unit of observation is a complete player's 3-round game history $H$. For instance, if $H=4$, we sampled the game histories of 4 random Blue players and computed the likelihood of this reduced dataset according to the models. We used a bootstrapping procedure to simulate the likelihood generated by datasets with fewer observations than the total actually collected in order to see the stability of these likelihoods. The final likelihood odds for each of the models is the value of the rightmost data point in each panel.}
    \label{fig:likelihoods3}
\end{figure}

\section{Behavioral Experiments}
We collected five datasets for five different experimental designs: (1) The design suggested by the modal expert response, (2) the optimal design determined by our algorithm, (3) a medium information gain design, (4) the lowest information design, and (5) the original experiment run by El-Gamal \& Palfrey (1996)  (see Fig.~\ref{fig:experimentsRun}B).

We recruited participants from the online labor market Amazon Mechanical Turk (AMT) \cite{horton2011online,mason2012conducting,paolacci2010running} and ran our experiments on the nodeGame platform for group experimentation online \cite{Balietti2016}, according to the experimental procedure outlined below.

\subsection{Experimental Protocol}
The experimental protocol resembled closely the description found in El-Gamal \& Palfrey (1996), with some adaptation for the online environment. All experiments were incentive-compatible and used no-deception; the protocol was approved by IRB (Northeastern IRB (\#13-13-09)).
 
As soon as an online worker accepted the task on AMT, he or she would receive the link to our experimental platform. Upon starting the experiment, each participant would first go through an instructions page and a brief tutorial alone, before being matched with other players in the interactive part. Our tutorial was recreated after the tutorial in the original El Gamal \& Palfrey (1996) paper, in which participants could choose to play a simulated game as either the Red or the Blue player.

After completing the tutorial, the participant would be moved into a waiting room. As soon as enough participants were present, a new game room would be launched. The waiting times in the waiting rooms were short (two to four minutes).

The interactive experiment would then proceed according to the rules explained in Sec.~\ref{sec:stopgo_rules}. Participants always had their cumulative score, the round count, and the time left for a decision visible at the top of the page; the history of own decisions from previous rounds was available upon clicking on the History button. More details, including the full instructions, how we handled dropouts, and screenshots of the interface, are available in the Sec.~\ref{sec:stopgo_exp_details} of the Appendix.

\subsection{Experiment Results}
In total, we collected behavioral data from 79 human players. The average duration of the interactive part of the experiment was 5 minutes and 20 seconds; the average payoff per treatment in US Dollars was: 1.15 ($A=2$ and $\pi=0.5$), 1.61 ($A=3.33$ and $\pi=0.5$), 1.97 ($A=6$ and $\pi=0.2$), 2.20 ($A=6$ and $\pi=0.5$), and 2.12 ($A=6$ and $\pi=0.8$).

Our goal was to collect as many experimental datasets as necessary to distinguish among the different models. Recall from Sec.~\ref{sec_oed_implementation} that in our experiment a dataset is composed by the joint game histories of a group of 10 players. After collecting just one dataset per experimental design, we were able to find out which model best described human behavior in our game. 

\begin{figure}[t!]
    \centering
    \includegraphics[width=1.0\columnwidth]{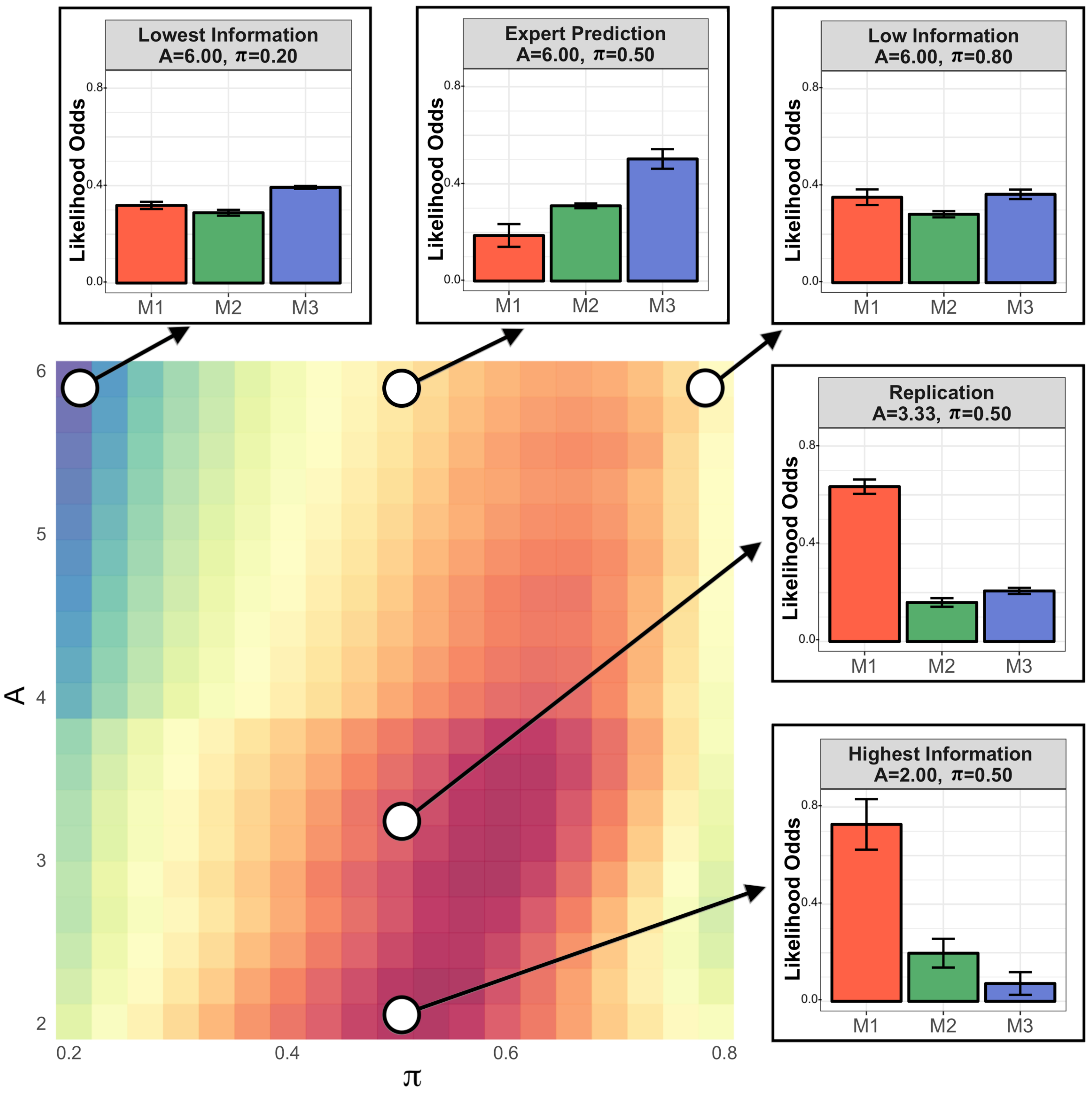}
    \caption{\textbf{Information surface with corresponding experimental results.} Intuitively, what we see here validates our expectation of the amount of information gain---the amount of divergence---between the likelihoods of the different models.}
    \label{fig:Experiments}
\end{figure}

The predicted likelihood of observing the data collected according to the three models are plotted as likelihood odds ratios in Fig.~\ref{fig:likelihoods3}. Heuristically, one can think of the plots where the likelihood odds are \textit{more different} as containing \textit{more information}. As such, it is encouraging to note that experimental designs (3.33, 0.5) and (2.0, 0.5) are the ones where the likelihoods of the different models diverge the most. These designs correspond to regions of high information in our information surface. We map the same data onto the information surface in Fig.~\ref{fig:Experiments}.

\section{Iterating the Procedure}
Across all experiments, we find strong evidence that Model 1 (the Bayes-Nash equilibrium prediction) is best suited to describe the observed data (i.e., Model 1 has the highest posterior odds). However, the process for selecting ``the best'' model never ends as we can never definitively accept one model as true \cite{El-Gamal1996EconomicalDesign}. The fact that one of the models performs well is not an indication that it will continue to do so when compared to other models based on new theories. When new theories of behavior are proposed, it is then desirable to test these new theories against the best known alternative. A key advantage of the optimal experimental design approach is that it can be iterated, allowing for sequential testing of new models. That is, the approach allows the comparison of predictions from \textit{new} models against the predictions of the most likely model currently available. 

We illustrate this important advantage of the optimal experimental design approach of iteratively expanding the set of behavioral models by testing a new theory that was not included in the original work on the imperfect information game: Reinforcement learning. That is, we repeated our optimal design protocol and included a fourth model that casts human behavior with a Roth-Erev reinforcement learning model \cite{Erev1998}.

Learning theories have become increasingly popular for describing human decision-making across a number of settings \cite{salmon2001evaluation,FoleyEtAl2018}. Recent reviews of the effectiveness of this model have shown that it is well-suited to describe human behavior in a number of settings, especially those where learning from experience is necessary in order to maximize a payout \cite{Erev2014}. Given promising results in other areas, Roth-Erev reinforcement learning proves to be a good candidate to study. Other models to compare could include those based on Experience-Weighted Attraction learning \cite{Camerer1999Experience-WeightedGames, Ho2008IndividualInformation}.

Once this new model is specified (i.e., it is able to assign a likelihood of observing a particular dataset given initial conditions), we are able to seamlessly incorporate it into our optimal experimental design procedure. The only change to this procedure comes when we calculate the information value associated with a given set of experimental design parameters; recall, this is done by calculating the KL divergence (Eq.~\ref{eq:DKL}) between the likelihoods (of observing a given dataset) generated by each of the different models. Whereas the values in the information surface from Figure \ref{fig:gpucbpe} are based on a KL divergence between three models, the values of this new information surface would be based on a comparison of four models. In SI Figure \ref{fig:four_model_surface}, we show this resulting surface, noting that the addition of the fourth model does not dramatically change the estimates about the optimal experiments.

\begin{figure}[t!]
    \centering
    \includegraphics[width=1.0\columnwidth]{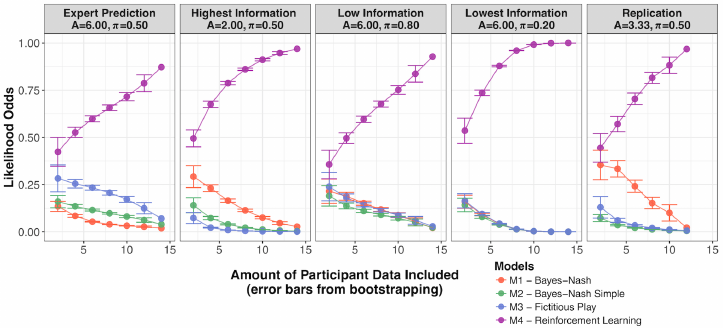}
    \caption{\textbf{Likelihood odd ratios of models according to real participants behavior including a fourth Model.} The unit of observation is a complete player's 3-round game history $H$. For instance, if $H=4$, we sampled the game histories of 4 random Blue players and computed the likelihood of this reduced dataset according to the models. After including Model 4---a reinforcement learning model---Model 1 is no longer the best-suited for describing this data.}
    \label{fig:odds4}
\end{figure}

When we include this fourth model in the analysis of empirical data, Model 1 is no longer the best model for describing the data (see Fig.~\ref{fig:odds4}). Reinforcement learning explains the data significantly better than any of the other models, across all five experimental designs. Through this process, we found that the new fourth model outperformed the other three models in every experiment that we ran. The reinforcement model reliably and robustly describes the data from five different experiment configurations better than the static equilibrium predictions. This also suggests that the reinforcement learning model is quite insensitive to the parameterization of the experiment, indicating its suitability for describing behavior. 

In summary, we make a substantive contribution showing that Roth-Erev reinforcement learning explains behavior in the imperfect information game better than competing models including fictitious play and Bayes-Nash equilibrium prediction. Observing such a strong fit in an experiment as simple as the Stop-Go game indicates that this model can (and should) be tested in new domains in order to assess its effectiveness across a number of settings. In a way, this finding is iconic and represents the fact that newer, better, more biologically and socially plausible models of behavior will emerge as technology and scientific understanding of human behavior co-evolve. Roth-Erev reinforcement learning was one such model that has proven to be effective in describing behavior.

\section{Discussion}\label{sec_discussion}
The goal of this paper was to investigate the notion of \textit{information gain} in experimentation to identify behavioral types in situations involving strategic interactions. Specifically, we sought to improve upon a classic optimal experimental design approach where the optimal experiment is defined as the experiment that \textit{maximizes the KL divergence} between the likelihoods of multiple competing models under a given experimental design. In addition to illustrating and improving upon existing methods, we asked whether domain experts would predict optimal experimental designs that corresponded to those generated algorithmically. Lastly, we illustrated the potential that optimizing experiments for information gain has by adding an additional model into the protocol. In a way, this highlights the \textit{adaptive} nature of optimizing experimental design for information gain---it allows for the evaluation of new models of behavior as they are  invented.

By implementing a modified version of the adaptive search GPUCB-PE algorithm, we were able to actively construct an information landscape---a surface that represents how much information gain is associated with each possible combination of experimental design parameters. In doing so, we also eliminated the computationally-costly task of computing model likelihoods for all possibly-observable datasets during the optimal design process. This was achieved by sampling model parameters and simulating likely datasets. While this process produced more noisy estimates of the information gain associated with each experimental design, this noise quickly disappears through repeated (GPUCB-PE) search on the information surface. In practical terms, this reduced the task of finding the optimal experiment from 72 hours of computation on a super-computing cluster to approximately an hour on a standard laptop, which suggests that this procedure could be run in real-time.

The information landscape we constructed using our Parameter-Sampled GPUCB-PE procedure informed us about which experimental designs to test in our replication of the Stop-Go game \cite{El-Gamal1996EconomicalDesign}. However, we supplemented this list of experimental designs with insight from another source: domain experts. We conducted a survey, reaching out to over 800 experts in behavioral economics and digital experimentation to assess whether their domain expertise could associate to optimal predictions about the most informative experiment to run. The expert prediction experiment showed that there was little consensus among the experts about which experimental design was optimal. Interestingly, the experts generally lacked confidence in their predictions of the optimal experimental design, and this also appeared to make them inclined to advocate for an experiment with \textit{higher} payouts to the participants. However in reality, the optimal experiment was predicted to be one with the \textit{smallest} payout.

Ultimately, we ran five experiments in an attempt to distinguish between competing models of behavior. We began our optimal design procedure with three competing models; after we collected data, we found that one model was well-equipped to explain data from several experiments, but it was \textit{not} the most likely model across every experiment we ran. For this reason, we chose to introduce a \textit{fourth} model in our model comparison---a simple reinforcement learning model. Model 4 ended up being the model that was best able to explain the observed data across each of the five experiments. This finding is key, and it demonstrates the potential benefits of an optimal design procedure like the one described here. It enables researchers to quickly introduce new competing models that can range from subtle tweaks to an existing model through to an entirely new model. All that is needed is to add the new model to the model comparison procedure, find the experimental design that is expected to be most informative, and evaluate it on existing data or collect a new batch of data to test the models (if the optimal experiment is different from data already collected). Given that it is unlikely that researchers will discover one \textit{best} model for describing human behavior, there is a constant drive towards improving upon existing models, exploring for novel ones, and actively comparing them via the scientific process. Our fourth model outperforms the initial three that were tested, illustrating this point. This also contributes to the growing body of work suggesting that reinforcement learning is good at describing human behavior \cite{feltovich2000reinforcement, gale1995learning, sarin2001predicting, fershtman2012dynamic} by adding experimental results for repeated imperfect information games.

The benefits of rapidly assigning expected information gain to different experimental designs have a very wide reach beyond the current applications of Parameter-Sampled GPUCB-PE. While the current approach is meant for distinguishing between generative models of human behavior and picking the right experiment to separate the models, it can be adapted to accommodate different goals in experimentation, like optimizing the number of participants, optimizing for effect size, or optimizing for monetary cost. Looking ahead, this is a step in the direction of optimizing social network experiments where the particular network topology is the experimental design to test. This is a notoriously challenging computational task \cite{Phan2015,Parker2017} that could be eased using the methods described here. 

Finally, our procedure can also be used by firms---especially digital companies like Google, Facebook, Twitter, and Microsoft that run thousands of experiments every month \cite{goldman2016experiments}---to identify the most advantageous combinations of parameters for products, user interfaces, and marketing strategies \cite{kohavi_experiments_survey_2009, schwartz_bandit_experiments_2017, berman_beyond_last_touch_2018}. Firms that successfully master experiment optimization can enormously reduce the costs of exploring less lucrative regions of the parameter space and achieve a durable competitive advantage over competitors \cite{kohavi_surprising_experiments_2017, Azevedo2018}.

\section{Conclusion}
Researchers develop models exemplifying behavioral types in order to understand the mechanisms and decision processes in situations of strategic interactions and information asymmetry. As such, the role of experimentation is paramount for model selection. That is, distinguishing and evaluating the effectiveness of different models for describing human behavior. Experimenters can now use the algorithmic tools described in this paper to rapidly determine the exact experimental design that would maximally distinguish between the competing models of behavior. As a proof of principle, we used this tool in a game of imperfect information under several different parameterizations.

In the future, it is not difficult to imagine the successful implementation of an \textit{automated} online experimentation platform built using the tools such those described in this work \cite{bakshy_ae_2018}. Such an experimental platform would assign the parameters of an experiment, run an optimal experiment, update the model predictions, and evaluate the differences between the competing behavioral models. This could be used to search high-dimensional information landscapes, comparing complex models with many parameters; such complexity would likely need informative priors assigned to the different model parameters in order for the Parameter-Sampled GPUCB-PE search to work best, and future work will explore such approaches. In a way, this places more emphasis on the scientist's role as a theoretician, integrating results from optimally-designed experiments into new models that capture the commonalities and peculiarities of human behavior. Together, they form a human-AI team that can help advance social science more rapidly.

\subsection*{Acknowledgements}
The authors acknowledge Mahmoud El-Gamal for helpful correspondences and Stephanie W. Wang for useful comments on the design and implementation. This work was supported in part by the Office of Naval Research (N00014-16-1-3005 and N00014-17-1-2542) and the National Defense Science \& Engineering Graduate Fellowship (NDSEG) Program.

\subsection*{Conflict of interest}
The authors declare that they have no conflict of interest.

\subsection*{Author contributions}
All authors contributed to the study conception and design. All authors contributed to analyses and preparation of the manuscript. S.B. ran the online experiments, and all authors contributed to the expert surveying.

\subsection*{Code and data availability}
The optimal experimental design code and the data for replicating the results in this study can be found at \url{http://github.com/shakty/optimal-design}.

\printbibliography[title={References}]
\vfill

\appendix
\renewcommand\thefigure{\thesection.\arabic{figure}}    

\section{Appendix}
\setcounter{figure}{0}    

\subsection{GPUCB-PE Algorithm}\label{appendix:GPUCBPE}
The Gaussian Process Upper Confidence Bound-Pure Exploration (GPUCB-PE) search algorithm is described herein. In our case, the algorithm iteratively searches different points on an information landscape where each point is a coordinate, a pair of parameter values $(A, \pi)$ that together define a single experimental design, $\theta$. The \textit{value} associated with any given point in this landscape is the information gain, $I(x, \theta)$, defined in Equation 2 in the main text.

\begin{figure}[H]
    \centering \includegraphics[width=1.0\columnwidth]{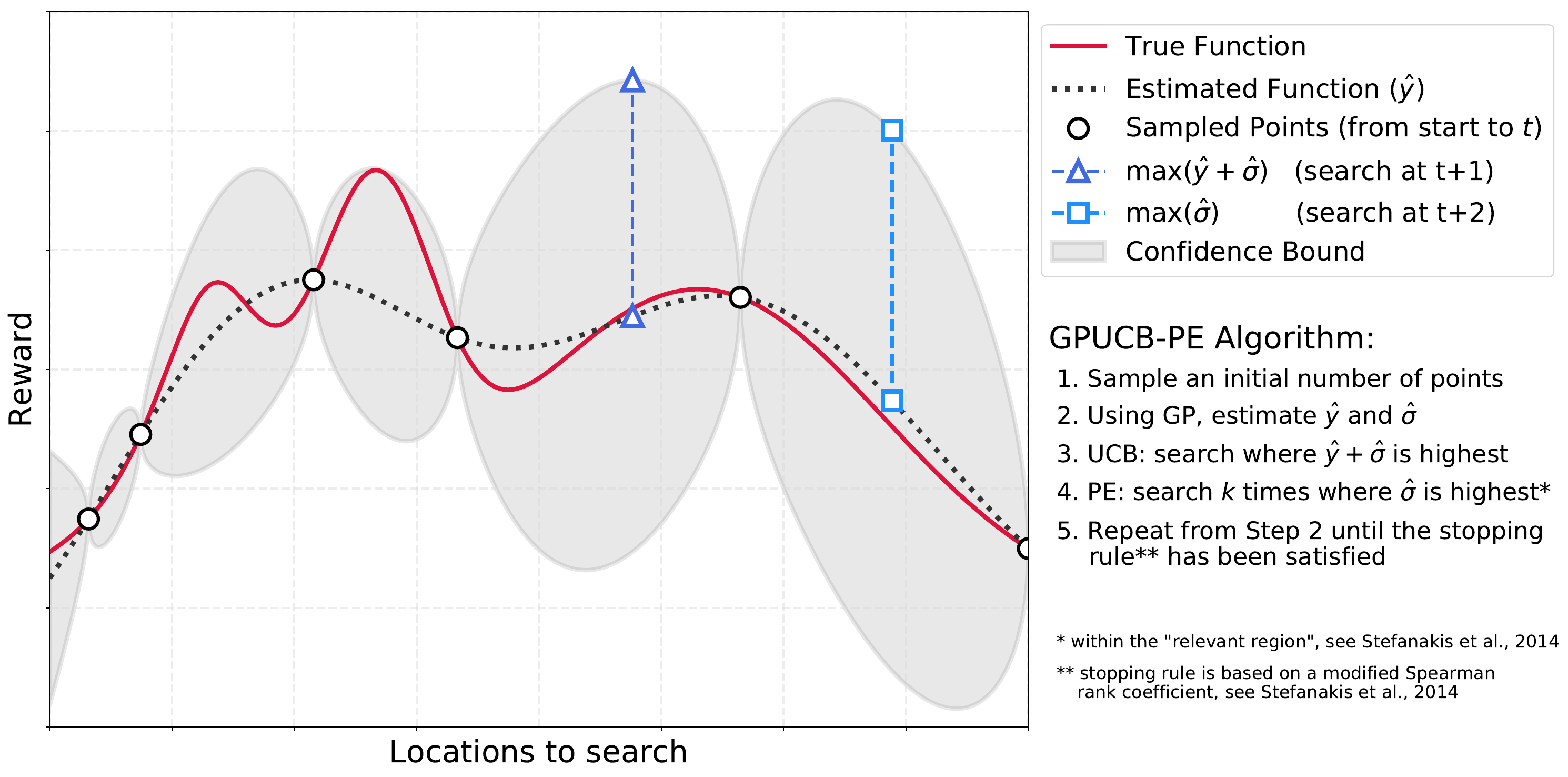}
    \caption{\textbf{Schematic representation of the GPUCB-PE algorithm.} 
    Gaussian Process regression has recently emerged as an effective tool for searching complex landscapes. As such, there are numerous software packages that simplify this task. In the current work, we rely on the GPfit package in the R statistical computing library \protect\cite{GPfit}, making the computations behind Step 2 (using GP, estimate $\hat{y}$ and $\hat{\sigma}$) straightforward.}
    \label{fig:GPUCB-PE}
\end{figure}

\begin{enumerate}
    \item Sample an initial number of points from the landscape and calculate the associated information gain, $I(x, \theta)$, of each.
    \item With these initial points, use Gaussian Processes to construct an estimate, $\hat{y}$, for the value of each other point in the landscape, along with the variance, $\hat{\sigma}$, associated with each $\hat{y}$ estimate.
    \item GPUCB-PE is associated with two types of search: a ``UCB'' search and a ``PE'' search. Each iteration of the algorithm performs one search at the coordinate associated with the highest $\hat{y} + UCB$ value, where $UCB$ is the upper confidence bound defined using the $\hat{\sigma}$ from the Gaussian Process in Step 2.
    \item The algorithm then selects \textit{k} ``pure exploration'' (PE) points to search (for more complex landscapes a higher $k$ may be needed, providing more exploration and making the algorithm more likely to converge to the global maximum). Each PE search selects a point in the landscape with the highest $\hat{\sigma}$ value within a ``relevant region'' so as to minimize the most uncertainty in the estimate of the landscape\footnote{Note: the ``relevant region'' is introduced so as to avoid searching points that have high uncertainty, $\hat{\sigma}$, but low value, $\hat{y}$. The region includes every point with a $\hat{y}+UCB$ value that is higher than the point with the largest $\hat{y}-LCB$ value, the largest \textit{lower} confidence bound). See \cite{Contal2013, Stefanakis2014CanApproach} for a more detailed description of the PE step in GPUCB-PE.}
    \item The algorithm then calculates the $I(x, \theta)$ of each newly sampled point and runs another Gaussian Process model, producing new estimates for $\hat{y}$ and $\hat{\sigma}$ that account for the addition of the new points.
    \item The algorithm continues from Step 3 unless the \textit{stopping rule} is satisfied. This stopping rule is based off of a modified Spearman rank correlation coefficient, comparing the relative values of each of the points in the \textit{previous} estimate of the landscape to the ranking of the same points in the \textit{current} estimate of the landscape. This measure quantifies the dissimilarity between subsequent estimates of the landscape. The GPUCB-PE algorithm will stop if that measure is less than the stopping parameter, $\rho$, for a number of iterations in a row, $L$. These two values are, in practice, set to be $L=4$ and $\rho=0.001$. This stopping rule ensures that the GPUCB-PE algorithm will not continue to search uninformative points on the landscape, which is usually associated with its convergence on the global maximum in the landscape.
\end{enumerate}

See Figure \ref{fig:GPUCB-PE}.

\subsubsection{Parameter-Sampled GPUCB-PE Regret}\label{sec:ps_gpucbpe_regret}

\begin{figure}[H]
    \centering \includegraphics[width=1.0\columnwidth]{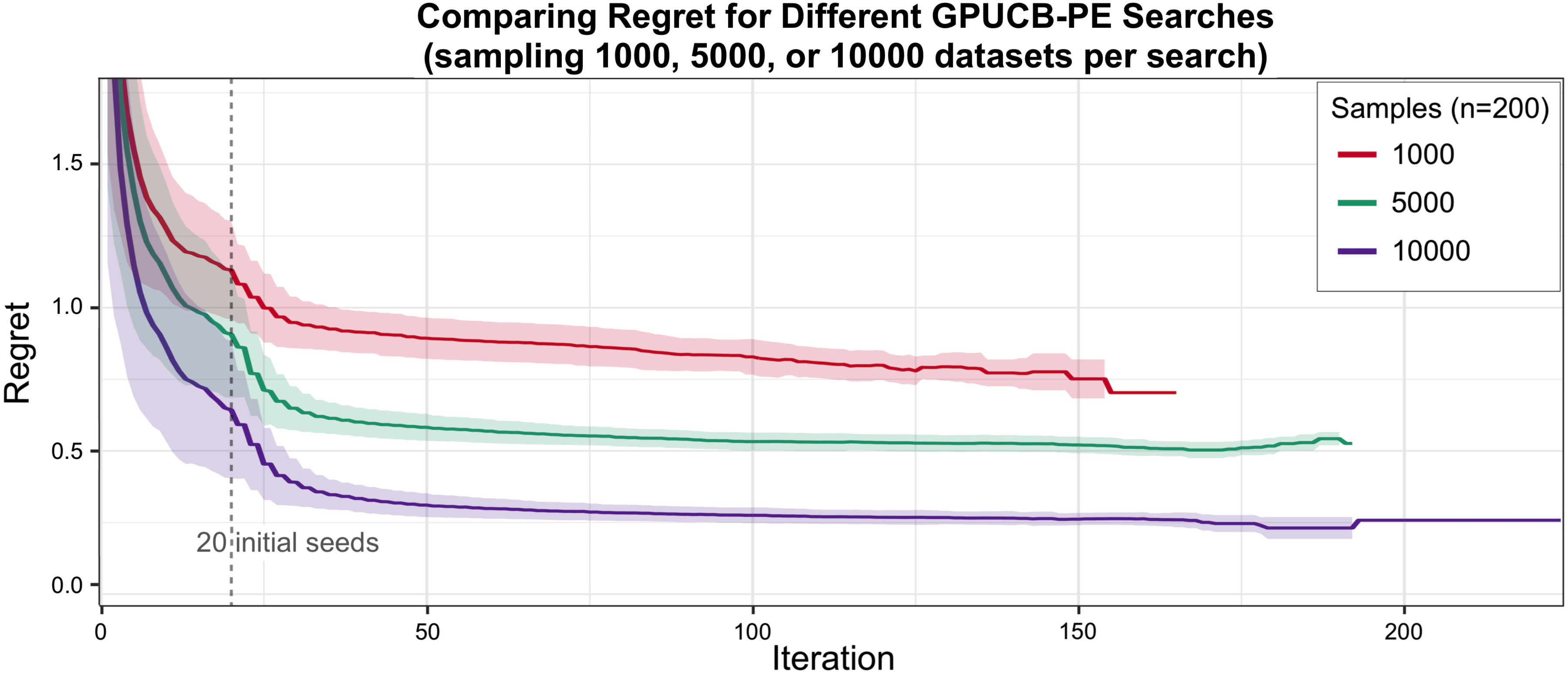}
    \caption{\textbf{Relationship between the number of sampled datasets per search in Parameter-Sampled GPUCB-PE (1000, 5000, or 10,000), the number of iterations needed to converge, and the regret (each simulated 200 times).} Sampling more datasets tends to require more searches in order for the algorithm to converge, while also further minimizing regret. Note: the differences in regret in this plot are due to the fact that sampling fewer datasets produces a relatively-higher mean squared error, but crucially, each of these simulations converged to the true global maximum ($\pi=0.5$ and $A=2.0$).
    The reason that these curves end up stabilizing at this optimal coordinate but different regret values is due to the nature of the information surface being searched and the differences in the distributions of likely datasets from the differently sized samples (1000, 5000, or 10000). If the variation in likely datasets did not differ greatly over these different conditions, the final regret values would be closer together. Conversely, with high variance in the distribution of likely datasets, we would expect these curves to be much farther apart.}
    \label{fig:regretGPcompare}
\end{figure}

\subsubsection{Parameter-Sampled GPUCB-PE Information Surface with Four Models}

\begin{figure}[H]
    \centering
    \includegraphics[width=1.0\columnwidth]{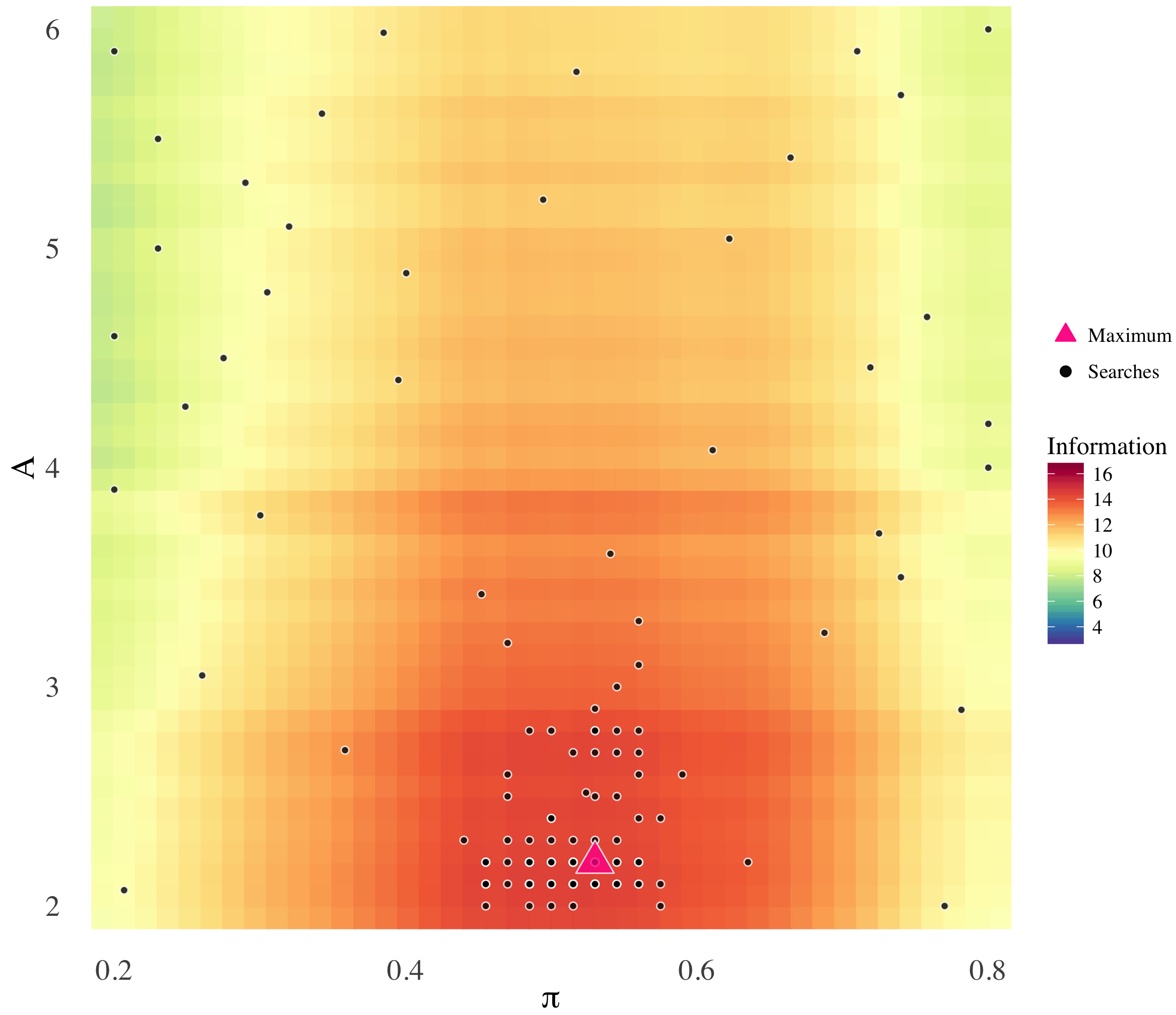}
    \caption{\textbf{Information surface generated by comparing four models.} By taking the average KL divergence between the likelihoods generated by each model we tested (including Model 4---the basic reinforcement learning model), we find an optimal experimental design that is consistent with the optimal experiment found in Figure 3, at approximately $A\approx 2$ and $\pi \approx 0.5$.}
    \label{fig:four_model_surface}
\end{figure}

\subsection{Expert Prediction Survey}
We implemented a Qualtrics survey for two weeks in early December 2017 in order to query the 788 experts in behavioral economics and digital experimentation. The survey was approved by Northeastern IRB (\#17-11-20) and tested before its launch in an internal pilot with researchers from Northeastern University. These researchers reported high confidence in the understanding of the questions. Below is the text used in the email recruitment for the survey and the survey itself.

\subsubsection{Recruitment Email Text}
\noindent Dear colleague:

\vskip 0.15in
\noindent We are researchers at the 
Network Science Institute at Northeastern University, and we would like to invite you to participate in a web-based online survey.

\vskip 0.15in
\noindent This study is about \textbf{experimental design}, and we are especially interested in how experts design experiments and make predictions about an experiment's outcomes. This survey is anonymous and is intended for academics in Economics and related disciplines. It is intended to take no more than 7 minutes.

\vskip 0.15in
\noindent Click here to start the survey.

\vskip 0.15in
\noindent On behalf of the Collaborative Social Systems Lab at Northeastern University, thank you very much for your time!

\vskip 0.15in
\noindent Sincerely,

\noindent Christoph Riedl, Assistant Professor and Principle Investigator

\noindent Stefano Balietti, Post-doc and Co-Principle Investigator

\noindent Brennan Klein, PhD Student and Research Assistant

\subsubsection{Survey Text}
\paragraph{Introduction:} Imagine you are a researcher who wants to study human behavior in a simple, two-player, repeated interaction game played over the course of three rounds. To do this, you decide to run an experiment that seeks to uncover a model that best describes human behavior. You believe there are four candidate models that may describe human behavior, and you are trying to find out which of them \textit{best} explains the behavior of participants in this game. We will ask you to choose values for two experimental parameters, \textbf{A} and \textbf{p}, for the game that you believe will \textit{maximally distinguish} the predictions of the four models. That is, you want to design the experiment such that it has the highest power to decide which model is most likely. We will then ask you to report which of the \textbf{four} models you think will best describe participants behavior in this game. First, we will ask you a few brief questions about your demographic information and education background\footnote{ Here, we asked participants to provide their current academic position, the country of their current academic institution, up to three research specialties, and their gender.}. Next, we will describe a simple incomplete information game played in pairs.

\vskip 0.15in
\paragraph{The Game:} In this game, there are two possible states of the world, \textit{a} and \textit{b}. The game starts with nature selecting state \textit{a} with probability \textbf{p} and state \textit{b} with probability \textbf{1 - p} (\textbf{p} is common knowledge). Player 1 is informed about the state of the world, while Player 2 is not. Player 1 then chooses to \textit{Stop} or \textit{Go}. \textit{Stop} ensures that both players receive a payout of 1, regardless of the state of the world. If \textit{Go} is selected, it is Player 2’s turn to select \textit{Left} or \textit{Right}. Finally, depending on the state of the world, Player 1 and Player 2 are given payouts according to those in Fig.~\ref{fig:stopgo}. This game is repeated for three rounds.

\vskip 0.15in
\noindent Your task is to choose values for the experimental parameters \textbf{A} and \textbf{p}.
\begin{itemize}
    \item \textbf{A} is the maximum payout that either player can receive
    \item \textbf{p} is the probability that the state of the world is \textit{a}
\end{itemize}
The values you choose for \textbf{A} and \textbf{p} should \textit{maximally distinguish} the predicted behavior of participants under each of the four models. That is, you want to pick values for \textbf{A} and \textbf{p} such that the \textit{predictions} generated by the four models will be \textit{most different}. On the following page, you will see descriptions of the four models.

\vskip 0.15in
\paragraph{The Models:}\footnote{ Note: because these models are rather conventional in the behavioral sciences, we did not include the full mathematical formulation of each model. Additionally, we were eliciting these experts' opinions and wanted to avoid them performing a full statistical analysis to arrive at their answers.}

\begin{itemize}
    \item \textbf{Model Q}: \textit{Each player plays the Bayes-Nash Equilibrium of the game.}
    In this model, each player predicts the other’s actions by forming beliefs about \textbf{p} (which determines the state of the world) and updating those beliefs after each round of play, based on the outcomes they have observed. 
    \item \textbf{Model R}: \textit{Similar to Model Q, but Player 2 does not update their beliefs about the state of the world.}
    In this model, each player predicts the other’s actions by forming beliefs about p (which determines the state of the world) and updating those beliefs after each round of play, based on the outcomes they have observed. 
    \item \textbf{Model S}: \textit{Each player engages in fictitious play to construct beliefs about their opponents’ actions.}
    In this model, each player keeps track of the state of the world and the actions taken by each player in every round they have played. They use this information to make predictions about \textbf{p} and what actions they should take in the current round. 
    \item \textbf{Model T}: \textit{Each player engages in Roth-Erev Reinforcement Learning.}
    In this model, a player is more likely to repeat an action if it had a positive outcome in the past. The propensity to select an action is a discounted sum of reinforcements obtained from playing the same action previously.     
\end{itemize}

\vskip 0.15in
\paragraph{Your Prediction:} Each of the four models assigns different likelihoods to the actions of Player 1 and Player 2. In order to distinguish between the four models, you will need to choose the experimental design where the \textit{predictions of the four models are most different}. This way, the data that you collect in your experiment will isolate one model as being more likely than the other three. 

\begin{itemize}
    \item What is your prediction for the best value of \textbf{A}? The values of \textbf{A} range from 2.0 $\leq \textbf{A} \leq$ 6.0.
    
    (text box)
    \item How confident are you in your prediction about \textbf{A}?
    
    (from 1 to 5)
    \item What is your prediction for the best value of \textbf{p}? The values of \textbf{p} range from 0.0 $< \textbf{A} <$ 1.0.
    
    (text box)
    \item How confident are you in your prediction about \textbf{p}?
    
    (from 1 to 5)
    \item Please rank the following four models based on which model do you think will best describe the participants’ behavior in the experiment you've proposed. Here, a rank of 1 is the most likely, and 4 is the least likely. NOTE: The names of the models (Q, R, S \& T) have been randomized and should not reflect the model's ranking at all.
    
    (dropdown, from 1 to 4)
    \item How confident are you in this ranking?
    
    (from 1 to 5)
    \item Optional: In the text box below, please provide 2-3 sentences about why you chose the model that you did.
    
    (text box)
\end{itemize}

\subsubsection{Survey Data}
\begin{table}[H]\small
\centering
\begin{tabular}{|| l | c | c ||} 
 \hline 
  \textbf{} & & \\
  \textbf{} & \textbf{Partial Completion} & \textbf{Full Completion} \\
  \textbf{} & & \\
 \hline\hline
 \textbf{Response Rate} & 0.18 & 0.07 \\
 \hline
 \textbf{Primary Field (select up to 3)} &  &  \\
 \hspace{.5cm}Behavioral Economics & 0.30 & 0.29 \\
 \hspace{.5cm}Game Theory & 0.17 & 0.16 \\
 \hspace{.5cm}Microeconomics & 0.12 & 0.10 \\
 \hspace{.5cm}Labor Economics & 0.07 & 0.08 \\
 \hspace{.5cm}Developmental Economics & 0.04 & 0.02 \\
 \hspace{.5cm}Information Economics & 0.04 & 0.04 \\
 \hspace{.5cm}Econometrics & 0.03 & 0.03 \\
 \hspace{.5cm}Financial Economics & 0.03 & 0.04 \\
 \hspace{.5cm}Business Economics & 0.03 & 0.03 \\
 \hspace{.5cm}Theoretical Economics & 0.02 & 0.01 \\
 \hspace{.5cm}Macroeconomics & 0.01 & 0.03 \\
 \hspace{.5cm}International Economics & 0.01 & 0.01 \\
 \hspace{.5cm}Other & 0.13 & 0.16 \\
 \hline
 \textbf{Academic Rank} & &  \\
 \hspace{.5cm}Full Professor & 0.09 & 0.03 \\
 \hspace{.5cm}Associate Professor & 0.16 & 0.05 \\
 \hspace{.5cm}Assistant Professor & 0.36 & 0.16 \\
 \hspace{.5cm}Lecturer & 0.03 & 0.01 \\
 \hspace{.5cm}Postdoctoral Researcher & 0.10 & 0.04 \\
 \hspace{.5cm}PhD Student & 0.14 & 0.06 \\
 \hspace{.5cm}Prefer not to say & 0.01 & 0.01 \\
 \hspace{.5cm}Other & 0.10 & 0.03 \\
 \hline
 \textbf{Gender} &  &  \\
 \hspace{.5cm}Male & 0.71 & 0.73 \\
 \hspace{.5cm}Female & 0.27 & 0.24 \\
 \hspace{.5cm}Prefer not to say & 0.02 & 0.03 \\
 \hspace{.5cm}Other & 0.00 & 0.00 \\
 \hline
 \textbf{Modal Parameter Value} &  &  \\
 \hspace{.5cm}$\pi$ & 0.50 & 0.50 \\
 \hspace{.5cm}$A$ & 6.00 & 6.00 \\
 \hline
 \textbf{Total Response Times} &  &  \\
 \hspace{.5cm}Median & -- & \hspace{0.2cm}10.93 min. \\
 \hspace{.5cm}Mean & -- & 105.02 min. \\
 \hspace{.5cm}Mean (removing $>$ 6hr.) & -- & \hspace{0.2cm}16.23 min. \\
 \hline
\end{tabular}
\caption{\textbf{Data about experts who responded to our survey (788 contacted).}}
\label{fig:experts}
\end{table}

The goal of the survey was to elicit the expert's wisdom about the best combination of A (i.e., payoff) and $\pi$ (i.e., likelihood of state of the world being \textit{a}) parameters; we find that a bit less than half of respondents converged to \textit{A}=6 and $\pi=0.5$ (25 out 55).

\subsubsection{Confidence, seniority and experts' estimates}

Herein, for the sake of completeness, we report additional findings about the relationship between experts' seniority, confidence, and their estimates.

\begin{figure}[t]
    \centering
    \begin{subfigure}{.326\textwidth}
        \centering
        \includegraphics[width=1.0\columnwidth]{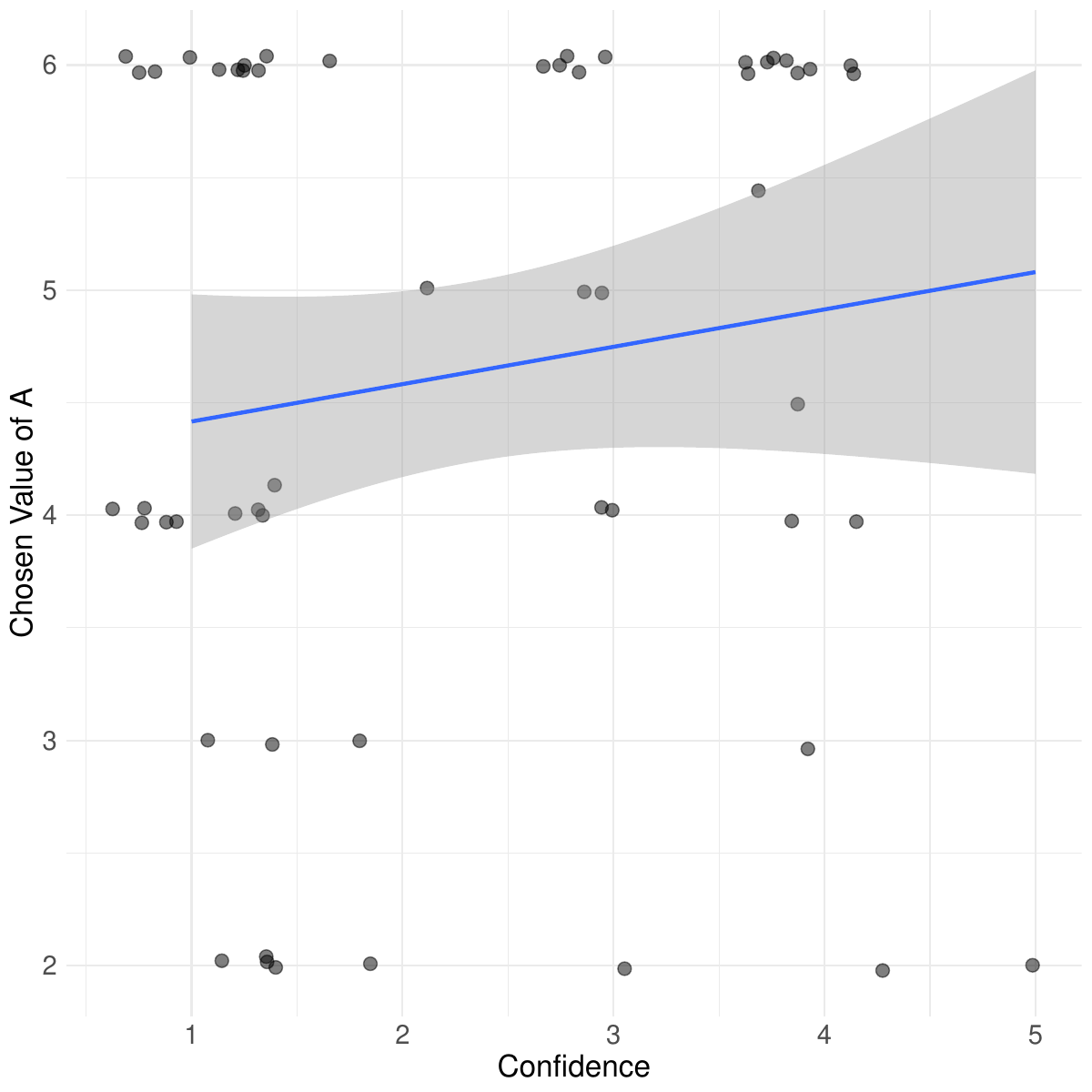}
        \caption{Confidence vs.~value of A.}
        \label{fig:conf_a}
    \end{subfigure}
    \begin{subfigure}{.326\textwidth}
        \centering
        \includegraphics[width=1.0\columnwidth]{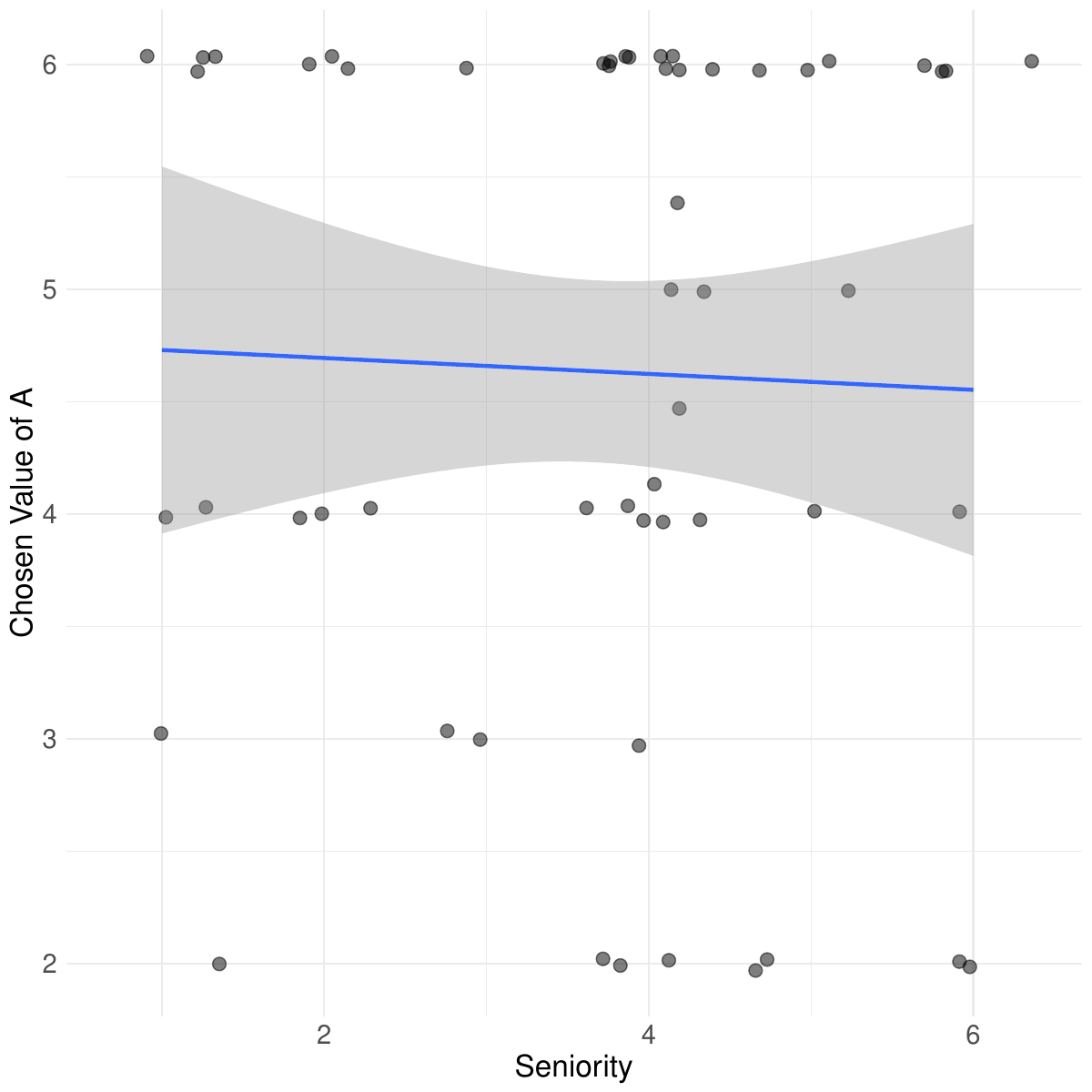}
        \caption{Seniority vs.~value of A.}
        \label{fig:seniority_a}
    \end{subfigure}
    \begin{subfigure}{.326\textwidth}
        \centering
        \includegraphics[width=1.0\columnwidth]{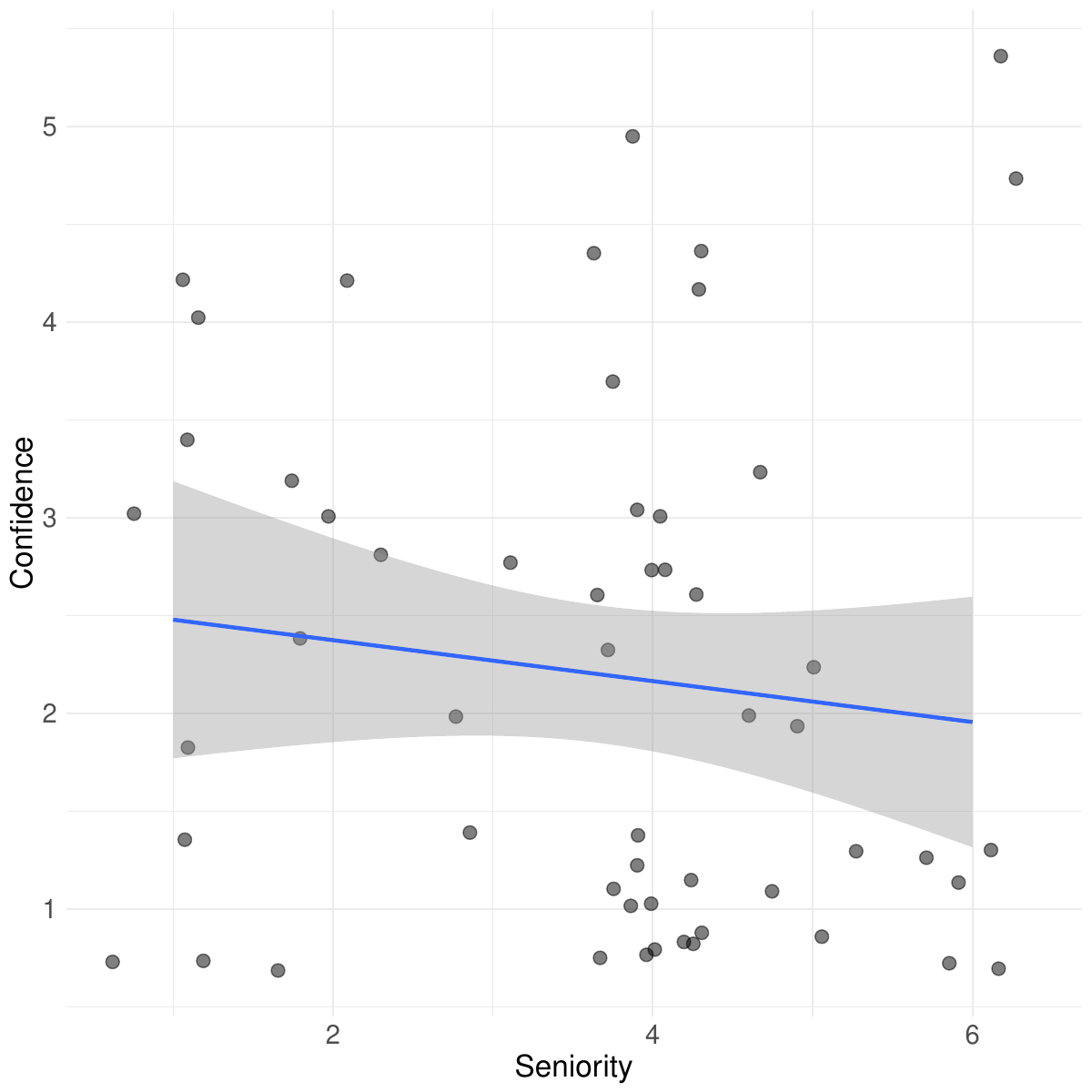}
        \caption{Seniority vs.~confidence in A.}
        \label{fig:seniority_confidence}
    \end{subfigure}
    \vskip\baselineskip
    \begin{subfigure}{.326\textwidth}
        \centering
        \includegraphics[width=1.0\columnwidth]{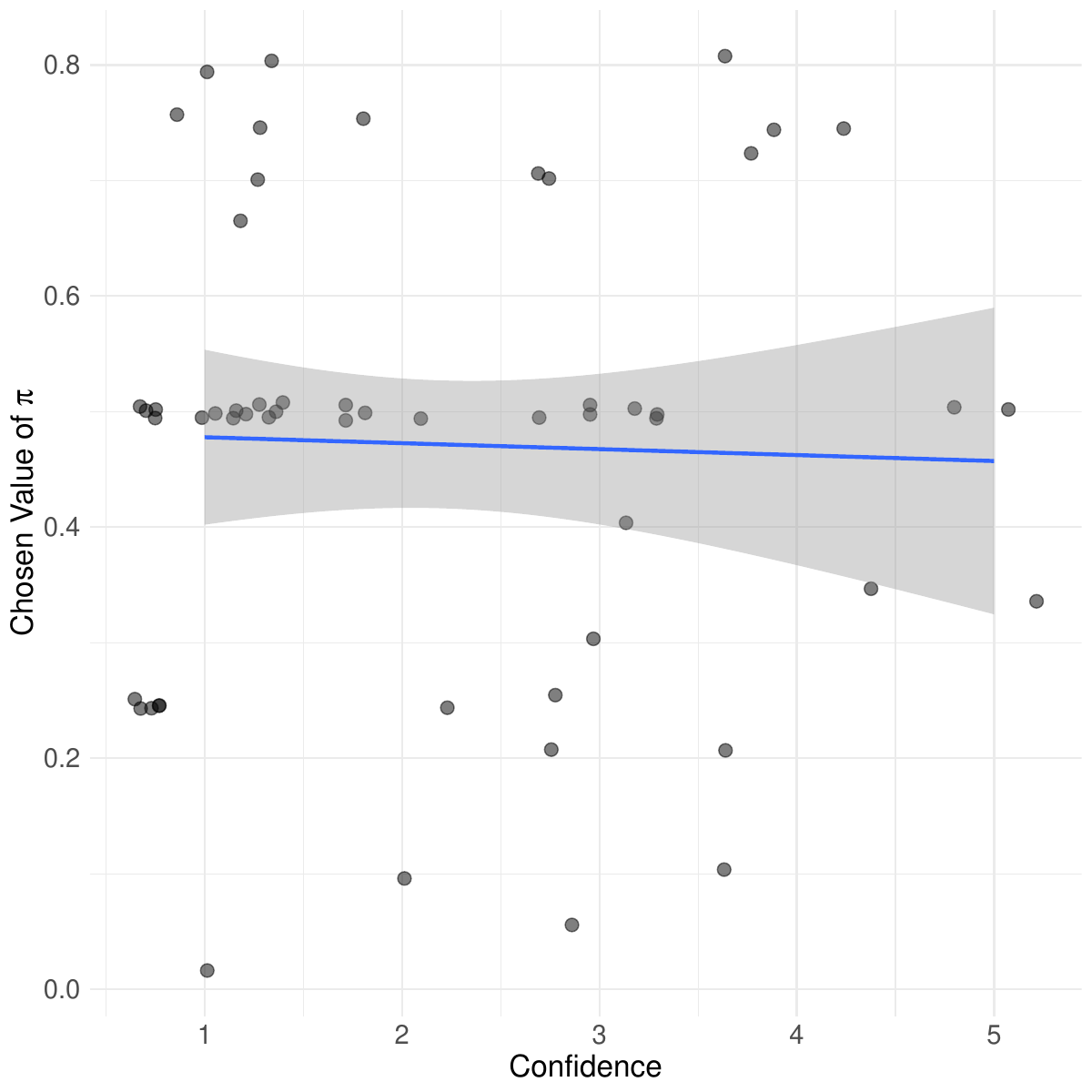}
        \caption{Confidence vs.~value of $\pi$.}
        \label{fig:conf_pi}
    \end{subfigure}
    \begin{subfigure}{.326\textwidth}
        \centering
        \includegraphics[width=1.0\columnwidth]{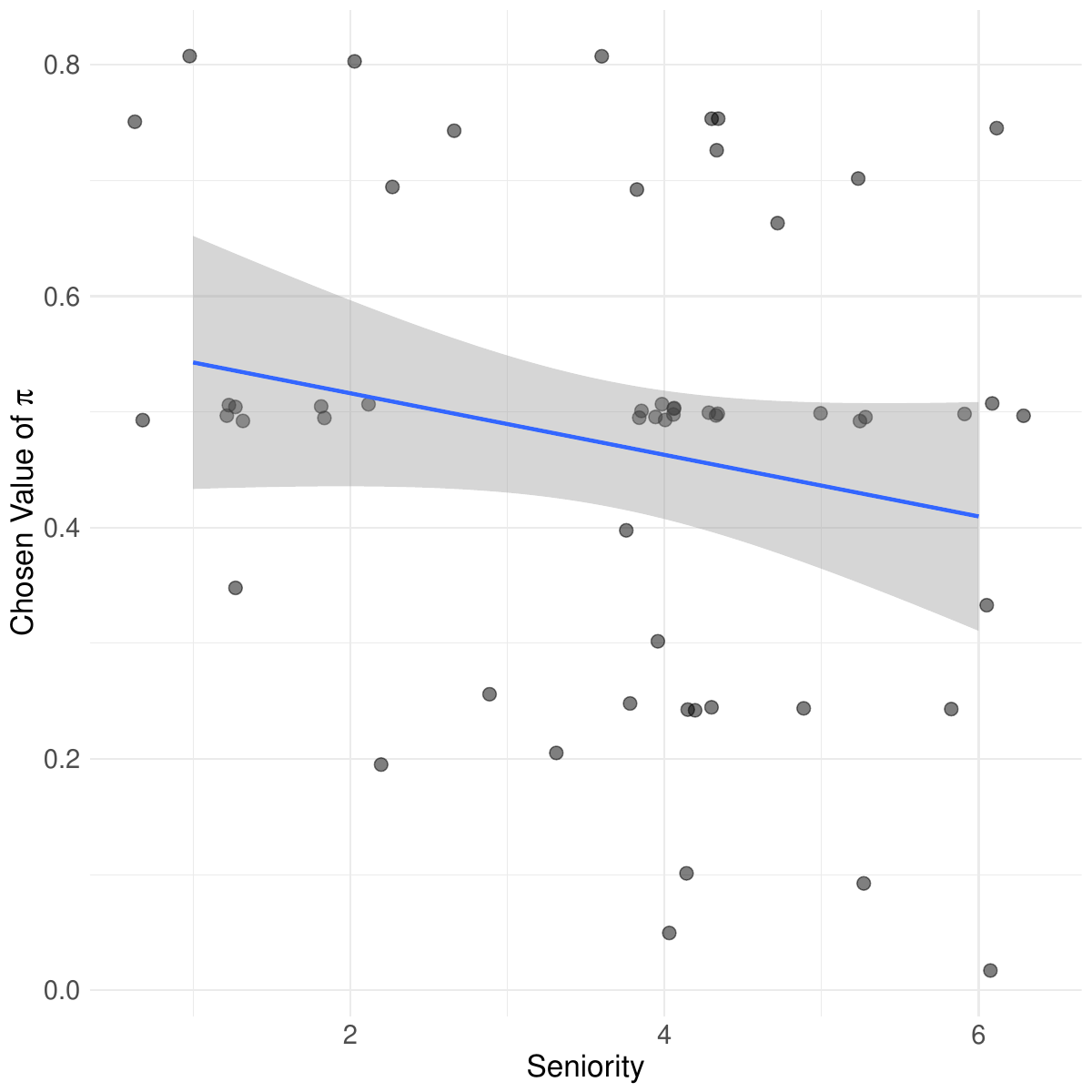}
        \caption{Seniority vs.~value of $\pi$.}
        \label{fig:seniority_pi}
    \end{subfigure}
    \begin{subfigure}{.326\textwidth}
        \centering
        \includegraphics[width=1.0\columnwidth]{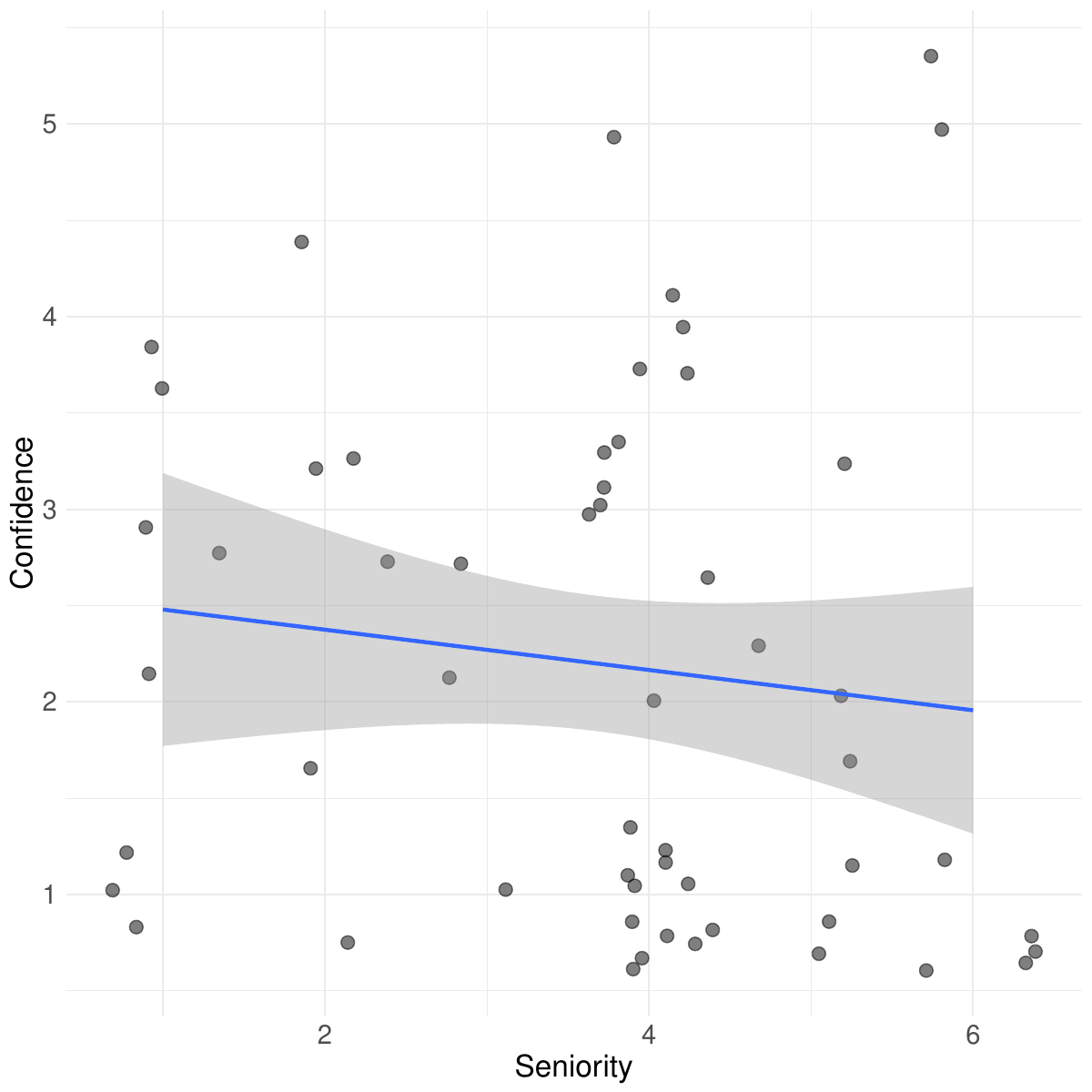}
        \caption{Seniority vs.~confidence in $\pi$.}
        \label{fig:seniority_confidence_pi}
    \end{subfigure}
    \caption{\textbf{Confidence, seniority and estimates of A (top row) and $\pi$ (bottom row).} All points are jittered; blue lines are linear regression fits and shaded areas are 95\% CI. Seniority is coded as follows: 1=PhD Student, 2=Postdoc, 3=Other, 4=Lecturer/Assistant Professor, 5=Associate Professor, 6=Full Professor (where information was available we manually coded the value of seniority for respondents who selected 'Other').}
\label{fig:seniority_confidence_API}
\end{figure}

Fig.~\ref{fig:seniority_confidence_API} shows these associations, none of which is statistically significant in our sample. Nonetheless, it is worth noting a couple of points: i) more senior researchers tend to be less confident in the estimates of \textit{A} and $\pi$, ii) for the estimate of parameter \textit{A}, where the experts disagree with the algorithm, we find a significant interaction effect between confidence and seniority. Seniority counteracts the upward pull of confidence on the estimate of \textit{A} in the direction chosen by the algorithm (see Table \ref{table:interaction_seniority}); however, this effect is not strong enough, and the resulting estimate of \textit{A} still far from the algorithmically optimal point.

\begin{table}[ht]
\begin{center}
\begin{tabular}{l c c c }
\hline
 & Model 1 & Model 2 & Model 3 \\
\hline
(Intercept)  & $4.36^{***}$ & $2.72^{**}$         & $2.32^{*}$          \\
             & $(0.65)$     & $(0.98)$            & $(1.13)$            \\
Confidence           & $0.16$       & $0.87^{*}$          & $0.85^{*}$          \\
             & $(0.15)$     & $(0.36)$            & $(0.36)$            \\
Seniority    & $-0.03$      & $0.41^{\texttt{+}}$ & $0.43^{\texttt{+}}$ \\
             & $(0.13)$     & $(0.24)$            & $(0.25)$            \\
Confidence {\footnotesize x} Seniority &              & $-0.19^{*}$         & $-0.19^{*}$         \\
             &              & $(0.09)$            & $(0.09)$            \\
Location US           &              &                     & $0.06$              \\
             &              &                     & $(0.45)$            \\
Gender Male   &              &                     & $0.39$              \\
             &              &                     & $(0.46)$            \\
\hline
R$^2$        & 0.02         & 0.11                & 0.12                \\
Adj. R$^2$   & -0.01        & 0.05                & 0.03                \\
Num. obs.    & 55           & 55                  & 55                  \\
RMSE         & 1.49         & 1.44                & 1.46                \\
\hline
\multicolumn{4}{l}{\scriptsize{$^{***}p<0.001$, $^{**}p<0.01$, $^*p<0.05$, $^{\texttt{+}}p<0.1$}}
\end{tabular}
\caption{\textbf{Linear regression predicting the value of A chosen by experts.}}
\label{table:interaction_seniority}
\end{center}
\end{table}

\begin{figure}[t]
    \centering
    \includegraphics[width=0.6\columnwidth]{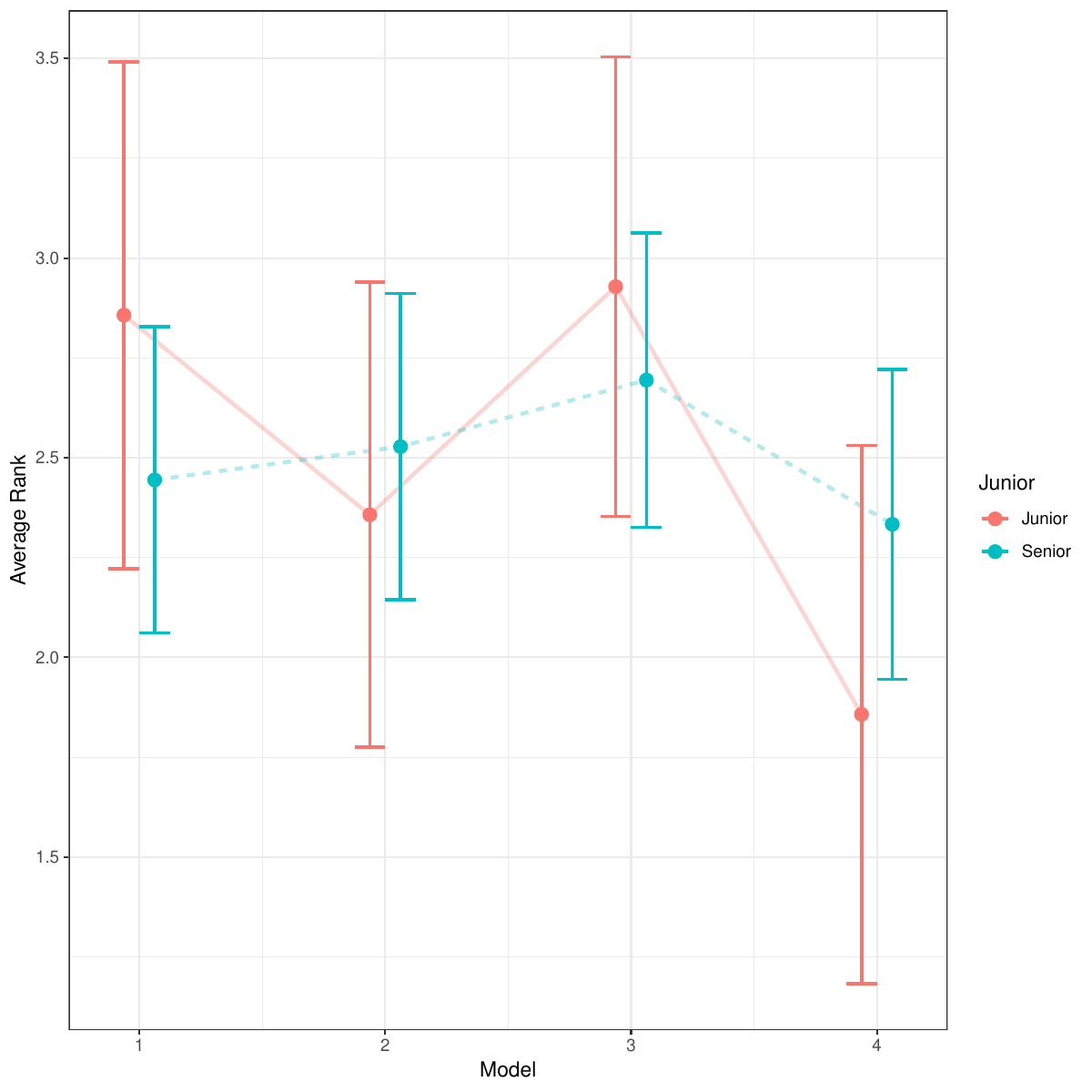}
    \caption{\textbf{Average model rankings by junior (PhD/postdoc) vs. senior researchers.} Lower values mean that a model is expected to fit the data better than others. Error bars: 95\% CI.}
    \label{fig:modelrank_junior}
\end{figure}

Finally, Fig.~\ref{fig:modelrank_junior} suggests a slight preference for the reinforcement model (Model 4) by more junior researchers, but the difference is not significant.

\subsubsection{Attrition and dropouts in the experts survey}\label{sec_survey_dropouts}

We performed statistical analysis to check possible differences between those who completed the survey and the rest of the invited sample. We modeled four logistic regressions predicting the binary outcome completion/not completion of the survey for four independent variables: gender (binary), region (North America, Europe, Asia, Other), position (7 levels, as reported above), and seniority (1-6 index, as reported above). 

We find no significant difference between people who started the survey (N=147) and completed it vs.~dropped out. When using the full invited sample (N=788), we still find no significant difference for gender and region. However, we find that more senior people are less likely to start the survey. Full details in Table \ref{table_dropout_survey_regr}.

\begin{table}
\begin{center}
\begin{tabular}{l c c | c c }
\hline
 & Start 1 & Start 2 & Finish 1 & Finish 2 \\
\hline
(Intercept)                     & $-0.74^{**}$  & $-0.51^{*}$   & $-0.09$  & $-0.48$  \\
                                & $(0.25)$      & $(0.25)$      & $(0.42)$ & $(0.46)$ \\
Seniority                       &               & $-0.24^{***}$ &          & $-0.01$  \\
                                &               & $(0.06)$      &          & $(0.12)$ \\
Postdoctoral Researcher & $0.67$        &               & $-0.32$  &          \\
                                & $(0.44)$      &               & $(0.67)$ &          \\
Other Position                   & $-1.43^{***}$ &               & $-1.01$  &          \\
                                & $(0.40)$      &               & $(0.79)$ &          \\
Lecturer                & $-2.15^{***}$ &               & $-1.01$  &          \\
                                & $(0.57)$      &               & $(1.23)$ &          \\
Assistant Professor     & $-0.25$       &               & $-0.21$  &          \\
                                & $(0.30)$      &               & $(0.50)$ &          \\
Associate Professor     & $-0.53$       &               & $-0.86$  &          \\
                                & $(0.34)$      &               & $(0.61)$ &          \\
Full Professor          & $-1.74^{***}$ &               & $-0.83$  &          \\
                                & $(0.38)$      &               & $(0.72)$ &          \\
\hline
AIC                             & 705.62        & 746.72        & 204.10   & 198.36   \\
BIC                             & 738.30        & 756.06        & 225.03   & 204.34   \\
Log Likelihood                  & -345.81       & -371.36       & -95.05   & -97.18   \\
Deviance                        & 691.62        & 742.72        & 190.10   & 194.36   \\
Num. obs.                       & 788           & 788           & 147      & 147      \\
\hline
\multicolumn{5}{l}{\scriptsize{$^{***}p<0.001$, $^{**}p<0.01$, $^*p<0.05$}}
\end{tabular}
\caption{\textbf{Logistic regression predicting start and completion of the survey by seniority and position.} ``Start'' models use the full sample and predict the odds of starting the survey; ``Finish'' models use only respondents who started the survey and predict the odds of completing it. Omitted regressions for Region and Gender, which are never significant.}
\label{table_dropout_survey_regr}
\end{center}
\end{table}

\subsection{Stop-Go Experiment}\label{sec:stopgo_exp_details}

We ran a total of 7 sessions for 5 experimental designs between Jan 18 and Jan 29 2018. We first ran the experimental design ``Replication'' ($A=3.33$ and $\pi=0.5$) with group size of 10 participants---as in the original El-Gamal \& Palfrey (1996) paper. However, this session experienced one dropout, which required us to re-run it. We then decided to start all next sessions with an overbooked group size of 14 participants, from which we could bootstrap groups of 10-player datasets. Even so, the experimental design ``Lowest Information'' ($A=6$ and $\pi=0.2$) experienced a dropout early on in the game, which required us re-run it.

In total, we only had two dropouts, a rate which is lower than in many other online group-behavior experiments \cite{Arechar2018, Stewart2017}. We cannot, therefore, draw any meaningful correlation with behavior and characteristics of dropouts, but we report here some additional information. Dropout 1 happened in the mid-paying treatment ($A=3.33$) and was a Red player who disconnected before taking any decision; Dropout 2 happened in a high-paying treatment ($A=6$) and was a Red player who disconnected in the first round, after taking a decision.

To ensure that our experiment provided equal earning opportunities for all players, we replaced dropouts with automated players (bots). This allows all players in a session to finish the experiment, even if another player got disconnected or dropped out. Bots replayed the decisions made by the majority of human players across all treatments in the same situation. However, for our analysis we discarded all data from human players who interacted with a bot, and from players who interacted with a player who interacted with a bot (and so on). Under these conditions, after collecting just one dataset per experimental design, we were able to find out which model best described human behavior in our game.

\subsection{Models of Behavior}\label{appendix:models}
We implemented the same models that were tested in El-Gamal \& Palfrey (1996):

\noindent\textit{Model 1}: Each individual, \textit{i}, plays the Bayes-Nash Equilibrium of the game, defined by ($\pi_{per}, A, \epsilon$).\\
\noindent\textit{Model 2}: Player 2 does not update $\pi_{per}$ following \textit{Go}, and this is common knowledge.\\
\noindent\textit{Model 3}: Individuals use fictitious play to construct beliefs about opponents' play.

\subsubsection{Model 1}
Model 1 is characterized by five canonical cases (\textit{A, B, C, D,} and \textit{E}) which correspond to particular combinations of parameters.

$\hat{\pi}_{per} = \dfrac{A}{2+A}$

\noindent\textbf{Case A}: $\pi_{per} > \hat{\pi}_{per}$ , $\epsilon \leq \dfrac{2\hat{\pi}_{per}(1-\pi_{per})}{\hat{\pi}_{per}+\pi_{per}-2\hat{\pi}_{per}\pi_{per}}$ , and $\epsilon \leq \dfrac{2}{A}$
\vspace{0.3cm}

$p_a^I = \dfrac{A(1-\pi_{per})}{2\pi_{per}} - \dfrac{[(A+2)\pi_{per}-A]\epsilon/2}{(1-\epsilon)2\pi_{per}}$

$p_b^I = 1$

$q^I = \dfrac{1}{1-\epsilon}\Big[ \dfrac{A-1}{A} - \epsilon/2 \Big]$

\vspace{0.3cm}
\noindent\textbf{Case B}: $\pi_{per} > \hat{\pi}_{per}$ , $\epsilon > \dfrac{2\hat{\pi}_{per}(1-\pi_{per})}{\hat{\pi}_{per}+\pi_{per}-2\hat{\pi}_{per}\pi_{per}}$ , and $\epsilon \leq \dfrac{2}{A}$
\vspace{0.3cm}

$p_a^I = 1 - p_b^I = 1 - q^I = 0$

\vspace{0.3cm}
\noindent\textbf{Case C}: $\pi_{per} \leq \hat{\pi}_{per}$ , $\epsilon \leq \dfrac{2\hat{\pi}_{per}(1-\pi_{per})}{\hat{\pi}_{per}+\pi_{per}-2\hat{\pi}_{per}\pi_{per}}$
\vspace{0.3cm}

$p_a^I = 1$

$p_b^I = \dfrac{2\pi_{per}}{A(1-\pi_{per})} + \dfrac{[(A+2)\pi_{per}-A]\epsilon/2}{(1-\epsilon)A(1-\pi_{per})}$ 

$q^I = \dfrac{1}{2}$

\vspace{0.3cm}
\noindent\textbf{Case D}: $\pi_{per} \leq \hat{\pi}_{per}$ , $\epsilon > \dfrac{2\pi_{per}(1-\hat{\pi}_{per})}{\hat{\pi}_{per}+\pi_{per}-2\hat{\pi}_{per}\pi_{per}}$
\vspace{0.3cm}

$p_a^I = 1 - p_b^I = 1 - q^I = 1$

\vspace{0.3cm}
\noindent\textbf{Case E}: $\pi_{per} > \hat{\pi}_{per}$
\vspace{0.3cm}

$p_a^I = p_b^I = q^I = 1$

\vspace{0.3cm}
\subsubsection{Model 2}
The behavior predicted under Model 2 is as follows:

\vspace{0.3cm}
\noindent\textbf{Case A}: $2\pi_{per} > A(1-\pi_{per})$

\vspace{0.3cm}
$p_a^{II}= 
\begin{cases}
    1      & \text{if} \hskip 1em \epsilon A / 2 > 1\\
    0.5    & \text{if} \hskip 1em \epsilon A / 2 = 1\\
    0      & \text{if} \hskip 1em \epsilon A / 2 < 1
\end{cases}$

\vspace{0.3cm}
$p_b^{II}=
\begin{cases}
    1      & \text{if} \hskip 1em 2(1-\epsilon/2) > 1\\
    0.5    & \text{if} \hskip 1em 2(1-\epsilon/2) = 1\\
    0      & \text{if} \hskip 1em 2(1-\epsilon/2) < 1
\end{cases}$

\vspace{0.3cm}
$q^{II} = 1$

\vspace{0.3cm}
\noindent\textbf{Case B}: $2\pi_{per} < A(1-\pi_{per})$

\vspace{0.3cm}
$p_a^{II}=
\begin{cases}
    1      & \text{if} \hskip 1em (1-\epsilon/2)A > 1\\
    0.5    & \text{if} \hskip 1em (1-\epsilon/2)A = 1\\
    0      & \text{if} \hskip 1em (1-\epsilon/2)A < 1
\end{cases}$

\vspace{0.3cm}
$p_b^{II} = q^{II} = 0$

\vspace{0.3cm}
\noindent\textbf{Case C}: $2\pi_{per} = A(1-\pi_{per})$

\vspace{0.3cm}
$p_a^{II} = 1.0$

\vspace{0.3cm}
$p_b^{II} = 0.5$

\vspace{0.3cm}
$q^{II} = 0.5$

\subsubsection{Model 3}
The behavior predicted under Model 3 is as follows:

\vspace{0.3cm}
$emp_t = \dfrac{emp + \{\text{num}_{left} | \text{num}_{go}\}}{\text{num}_{go}+1}$ 

\vspace{0.3cm}
$empa_t = \dfrac{empa + \{\text{num}_{go} | \text{num}_{game_a}\}}{\text{num}_{go}+1}$ 

\vspace{0.3cm}
$empb_t = \dfrac{empb + \{\text{num}_{go} | \text{num}_{game_b}\}}{\text{num}_{go}+1}$ 

\vspace{0.3cm}
\noindent Making Player 2's updated belief at round \textit{t} equivalent to  $emp\pi_t$ below:

\vspace{0.3cm}
$emp\pi_t = \dfrac{empa_t\pi_{per}}{empa_t\pi_{per}+empb_t\pi_{per}}$

\vspace{0.3cm}
\noindent Such that:

\vspace{0.3cm}
$p_{a,t}^{III}=
\begin{cases}
    1      & \text{if} \hskip 1em (1-emp_t)A > 1\\
    0.5    & \text{if} \hskip 1em (1-emp_t)A = 1\\
    0      & \text{if} \hskip 1em (1-emp_t)A < 1
\end{cases}$

\vspace{0.3cm}
$p_{b,t}^{III}=
\begin{cases}
    1      & \text{if} \hskip 1em 2 emp_t > 1\\
    0.5    & \text{if} \hskip 1em 2 emp_t = 1\\
    0      & \text{if} \hskip 1em 2 emp_t < 1
\end{cases}$

\vspace{0.3cm}
$q_t^{III}=
\begin{cases}
    1      & \text{if} \hskip 1em 2 emp\pi_t > (1-emp\pi_t)A\\
    0.5    & \text{if} \hskip 1em 2 emp\pi_t = (1-emp\pi_t)A\\
    0      & \text{if} \hskip 1em 2 emp\pi_t < (1-emp\pi_t)A
\end{cases}$

\subsubsection{Calculating the Likelihood of Every Possible Dataset}
The likelihood of a particular \textit{move} for a player is defined as: 

\vspace{0.3cm}
$obsp_{a,t}^M = (1-\epsilon_t)p_{a,t}^M+\epsilon_t/2$

\vspace{0.3cm}
$obsp_{b,t}^M = (1-\epsilon_t)p_{b,t}^M+\epsilon_t/2$

\vspace{0.3cm}
$obsq_{t}^M = (1-\epsilon_t)q_{t}^M+\epsilon_t/2$

\vspace{0.3cm}
\noindent Using these values, the likelihood for a particular \textit{model}, $M \in \{I,II,III\}$, is defined by:
\begin{align}
like_{T}^M = \int\limits_{\epsilon_t} \int\limits_{\alpha} \int\limits_{\delta} \int\limits_{\pi-\delta}^{\pi+\delta} \Big( \prod\limits_{t=1}^{T} \prod\limits_{i=1}^{n} like(action_{t}^i | M; \epsilon_0, \alpha, \pi_{per}\Big) d\pi_{per}prior(d\epsilon_0, d\alpha, d\delta)
\label{eq:likelihood}
\end{align}

\noindent Using these likelihood functions in equation \ref{eq:likelihood}, the long process of computing all possible datasets begins. For each coordinate in a discrete grid of A and $\pi$---from [2, 6] and [0.2, 0.8] respectively---we compute the likelihood of each Model (1, 2, 3) at that coordinate. This, in essence, allows us to see which design parameters will optimally distinguish our competing models. 

\subsubsection{Screenshots of Instructions and Interface}

\begin{figure}[ht]
    \centering
    \includegraphics[width=0.85\columnwidth]{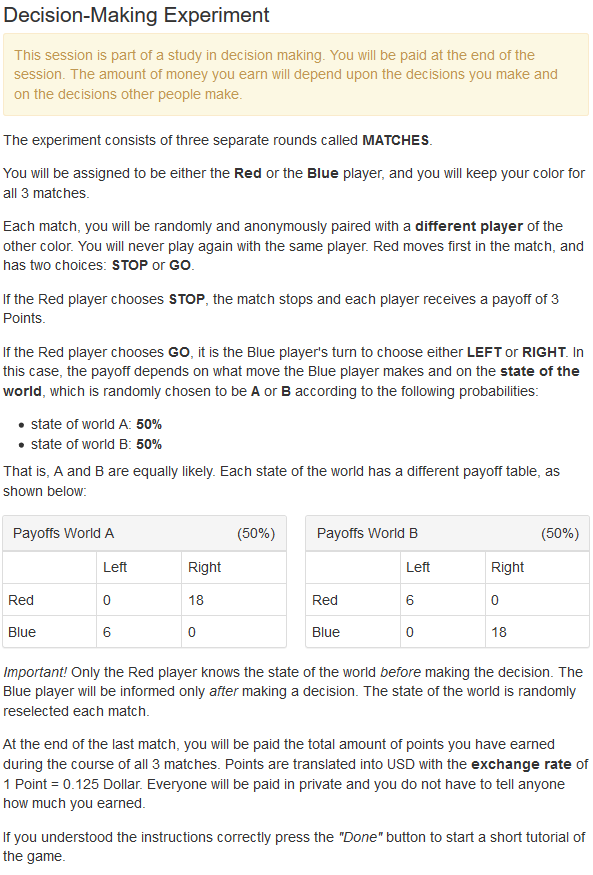}
    \caption{Instructions for $A=6.0, \pi=0.50$.}
    \label{fig:stopgo_instructions}
\end{figure}

\begin{figure}[ht]
\centering
\begin{subfigure}{1\textwidth}
  \centering
  \includegraphics[width=0.85\columnwidth]{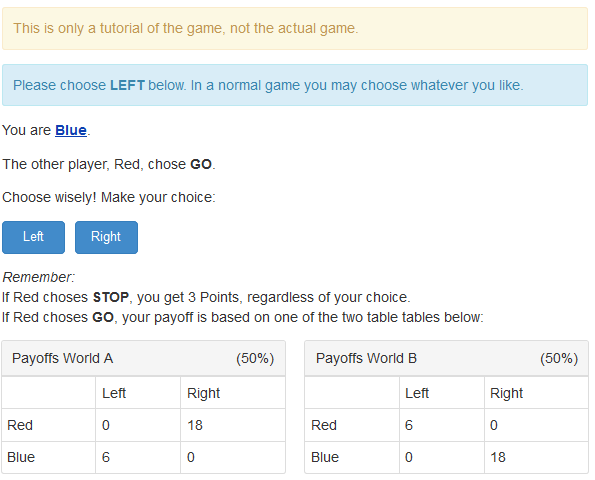}
  \caption{A decision step of the tutorial of the Blue Player.}
  \label{fig:stop_tutorial_blue}
\end{subfigure}
\begin{subfigure}{1\textwidth}
  \centering
  \includegraphics[width=0.85\columnwidth]{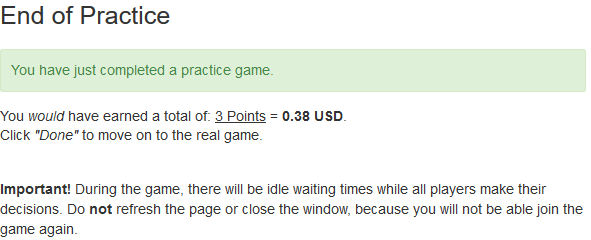}
  \caption{The end of the tutorial.}
  \label{fig:stopgo_tutorial_end}
\end{subfigure}
  \caption{Screenshots from the tutorial.}
  \label{fig:stopgo_tutorial}
\end{figure}

\begin{figure}[ht]
    \centering
    \includegraphics[width=0.85\columnwidth]{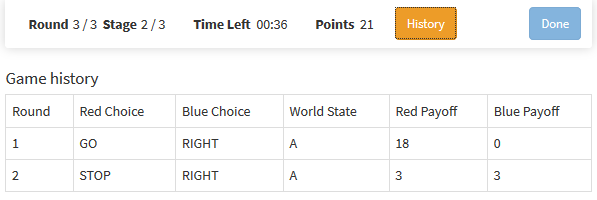}
    \caption{Game header: timer, earning, stage, and history of previous rounds.}
    \label{fig:stopgo_header}
\end{figure}

Herein, we show screenshots of the full instructions (Fig.~\ref{fig:stopgo_instructions}), the tutorial (Fig.~\ref{fig:stopgo_tutorial}), and the game header with information about cumulative score, progress in the game, time left for a decision, and the history of past decisions (Fig. \ref{fig:stopgo_header}).

\end{document}